\documentclass[prd,a4paper,superscriptaddress,nofootinbib]{revtex4}
\usepackage{amsmath,amssymb,epsfig,natbib,bm}

\begin{document}


\newcommand{\link}[1]{#1}

\newcommand{\edth}{\,\eth\,}
\renewcommand{\beth}{\,\overline{\eth}\,}

\newcommand{\deth}{{}_1\text{d}l\:}
\newcommand{\dbeth}{{}_1\overline{\text{d} l}\:}

\newcommand{\begm}{\begin{pmatrix}}
\newcommand{\enm}{\end{pmatrix}}

\newcommand{\Tr}{\text{tr}}
\newcommand{\baredth}{\beth}
\newcommand{\grad}{\nabla}

\newcommand{\lmax}{l_{\text{max}}}
\newcommand{\Bt}{\tilde{B}}
\newcommand{\Et}{\tilde{E}}
\newcommand{\bld}[1]{\mathrm{#1}}
\newcommand{\mC}{\bm{C}}
\newcommand{\mQ}{\bm{Q}}
\newcommand{\mU}{\bm{U}}
\newcommand{\mX}{\bm{X}}
\newcommand{\mV}{\bm{V}}
\newcommand{\mP}{\bm{P}}
\newcommand{\mR}{\bm{R}}
\newcommand{\mW}{\bm{W}}
\newcommand{\mD}{\bm{D}}
\newcommand{\mI}{\bm{I}}
\newcommand{\mM}{\bm{M}}
\newcommand{\mN}{\bm{N}}
\newcommand{\mS}{\bm{S}}
\newcommand{\mzero}{\bm{0}}

\newcommand{\btheta}{\bm{\theta}}
\newcommand{\bphi}{\bm{\phi}}

\newcommand{\vA}{\mathbf{A}}
\newcommand{\vAt}{\tilde{\mathbf{A}}}
\newcommand{\ve}{\mathbf{e}}
\newcommand{\vE}{\mathbf{E}}
\newcommand{\vB}{\mathbf{B}}
\newcommand{\vEt}{\tilde{\mathbf{E}}}
\newcommand{\vBt}{\tilde{\mathbf{B}}}
\newcommand{\vEw}{\mathbf{E}_W}
\newcommand{\vBw}{\mathbf{B}_W}
\newcommand{\vx}{\mathbf{x}}
\newcommand{\vX}{\mathbf{X}}
\newcommand{\vY}{\mathbf{Y}}
\newcommand{\vBwr}{{\vBw^{(R)}}}
\newcommand{\RW}{{W^{(R)}}}

\newcommand{\mUt}{\tilde{\mU}}
\newcommand{\mVt}{\tilde{\mV}}
\newcommand{\mDt}{\tilde{\mD}}

\newcommand{\rhat}{\hat{r}}
\newcommand{\half}{{\textstyle \frac{1}{2}}}
\newcommand{\third}{{\textstyle \frac{1}{3}}}
\newcommand{\numfrac}[2]{{\textstyle \frac{#1}{#2}}}
\newcommand{\ra}{\rangle}
\newcommand{\la}{\langle}
\newcommand{\qoe}{\numfrac{q}{\epsilon}}
\newcommand{\dlfdlq}{\frac{d \ln F}{d \ln q}}
\newcommand{\ixb}{{(b)}}
\newcommand{\ixnu}{{(\nu)}}
\newcommand{\ixgam}{{(\gamma)}}
\newcommand{\ixc}{{(c)}}
\newcommand{\lm}{{(lm)}}
\newcommand{\ginv}{{(\text{gi})}}
\newcommand{\upi}{\ubar{\pi}}
\newcommand{\usig}{\ubar{\sigma}}
\newcommand{\vort}{\varpi}

\newcommand{\cld}{\nabla}

\newcommand{\cla}{\mathcal{A}}
\newcommand{\clb}{\mathcal{B}}
\newcommand{\clc}{\mathcal{C}}

\newcommand{\cle}{\mathcal{E}}
\newcommand{\clf}{\mathcal{F}}
\newcommand{\clg}{\mathcal{G}}
\newcommand{\clh}{\mathcal{H}}
\newcommand{\cli}{\mathcal{I}}
\newcommand{\clj}{\mathcal{J}}
\newcommand{\clk}{\mathcal{K}}
\newcommand{\cll}{\mathcal{L}}
\newcommand{\clm}{\mathcal{M}}
\newcommand{\cln}{\mathcal{N}}
\newcommand{\clo}{\mathcal{O}}
\newcommand{\clp}{\mathcal{P}}
\newcommand{\clq}{\mathcal{Q}}
\newcommand{\clr}{\mathcal{R}}
\newcommand{\cls}{\mathcal{S}}
\newcommand{\clt}{\mathcal{T}}
\newcommand{\clu}{\mathcal{U}}
\newcommand{\clv}{\mathcal{V}}
\newcommand{\clw}{\mathcal{W}}
\newcommand{\clx}{\mathcal{X}}
\newcommand{\cly}{\mathcal{Y}}
\newcommand{\clz}{\mathcal{Z}}
\renewcommand{\H}{\clh}
\newcommand{\sigthom}{\sigma_T}
\newcommand{\Al}{{A_l}}
\newcommand{\Bl}{{B_l}}
\newcommand{\eAl}{e^\Al}
\newcommand{\ix}{{(i)}}
\newcommand{\ixp}{{(i+1)}}
\renewcommand{\k}{\beta}

\newcommand{\HD}{\mathrm{D}}
\newcommand{\curl}{\mathrm{curl}}
\newcommand{\CMBFAST}{\textsc{CMBFAST}}
\newcommand{\Omtot}{\Omega_{\mathrm{tot}}}
\newcommand{\Omb}{\Omega_{\mathrm{b}}}
\newcommand{\Omc}{\Omega_{\mathrm{c}}}
\newcommand{\Omm}{\Omega_{\mathrm{m}}}
\newcommand{\Oml}{\Omega_\Lambda}
\newcommand{\OmK}{\Omega_K}
\newcommand{\Ym}{{}_{-2}Y_{lm}}

\title{Analysis of CMB polarization on an incomplete sky}

\author{Antony Lewis}
 \email{Antony@AntonyLewis.com}
 \affiliation{DAMTP, CMS, Wilberforce Road, Cambridge CB3 0WA, UK.}
\author{Anthony Challinor}
 \email{A.D.Challinor@mrao.cam.ac.uk}
 \affiliation{Astrophysics Group, Cavendish Laboratory, Madingley Road, Cambridge CB3 OHE, UK.}
\author{Neil Turok}
 \email{N.G.Turok@damtp.cam.ac.uk}
 \affiliation{DAMTP, CMS, Wilberforce Road, Cambridge CB3 0WA, UK.}

\begin{abstract}
\vspace{\baselineskip}
The full sky cosmic microwave background polarization field can be
decomposed into `electric' and `magnetic' components.  Working in
harmonic space we construct window functions that allow clean separation
of the electric and magnetic modes from
observations over only a portion of the sky.
We explicitly demonstrate the method for
azimuthally symmetric patches, but also present it in a form in
principle applicable to
arbitrarily-shaped patches. From the window functions we obtain
variables that allow for robust estimation of the magnetic component
without risk of contamination from the probably much larger electric
signal. The variables have a very simple noise properties, and further analysis using them should be no harder than
analysing the temperature field. For an azimuthally-symmetric patch,
such as that obtained from survey missions when the galactic region is removed,
the exactly-separated variables are fast to compute. We 
estimate the magnetic signal that could be detected by the Planck
satellite in the absence of extra-galactic foregrounds. 
We also discuss the
sensitivity of future experiments to tensor modes in the presence of a
magnetic signal generated by weak lensing, and give lossless methods for analysing the
electric polarization field in the case that the magnetic component is negligible.
A series of appendices review the spin weight formalism and give recursion
relations  for fast computation of the spin-weighted spherical harmonics and
their inner products over azimuthally-symmetric patches of the sphere.
A further appendix discusses the statistics of weak signal detection.

\end{abstract}

\pacs{???}

\maketitle
\section{Introduction}

Observations of fluctuations in the temperature of the cosmic
microwave background (CMB) are now providing us with a 
direct view of the primordial inhomogeneities in the universe. 
The power spectrum of temperature fluctuations yields 
a wealth of information on the
nature of the primordial perturbations, and the values of the
cosmological parameters. Mapping the polarization of
the cosmic microwave sky is an important next step, offering 
a great deal of complementary information, especially
regarding the character of the primordial inhomogeneities~\cite{Bucher01}.
One of the
most interesting questions to resolve
is whether the primordial perturbations possessed a 
tensor (gravitational wave) component,
as predicted by simple inflationary models. Here, polarization 
measurements 
offer a unique probe~\cite{Kamionkowski97,Zaldarriaga97,Hu98}. 

Polarization of the cosmic microwave sky is produced by electron
scattering, as photons decouple from the primordial plasma. 
Linear polarization is produced when there is a quadrupole
component to the flux of photons incident on a scattering
electron. Scalar (density) 
perturbations generate an `electric' (gradient) polarization pattern
on the sky due to 
gradients in the velocity field on the surface of last scattering.
For scalar perturbations the velocity field is curl-free, and this leads
directly to the production of an entirely `electric' pattern
of linear polarization. In contrast, tensor perturbations 
(gravitational waves) produce polarization
by anisotropic redshifting of the energy of photons through decoupling.
In this case the polarization has `magnetic' (i.e.\ curl) and `electric'
(i.e.\ gradient) components at a comparable level. A magnetic
signal can also be produced by weak lensing of the electric
polarization generated by scalar modes. Detection and analysis of the
lensing signal would be interesting in itself, but a detection of an additional
tensor component would provide strong evidence
for the presence of primordial gravitational waves, a 
generic signature of simple inflationary models.  

Detecting or excluding a magnetic component is clearly of
fundamental significance in cosmology. But there is a 
significant obstacle to be faced. The problem is
that for the foreseeable future, the primordial sky
polarization will only be observable over the region
of the sky which is not contaminated by emission from our galaxy
and other foreground sources of polarization. Thus we
shall only be able to measure the polarization over a fraction of
the sky.
But the electric/magnetic decomposition is inherently \emph{non-local},
and \emph{non-unique} in the presence of boundaries. 

To understand this, consider the analogous problem of representing 
a vector field 
$V_i$ (in two dimensions) as a gradient plus 
a curl:
\begin{equation}
V_i = \nabla_i \Phi + \epsilon_{i}^{\,\,\,j} \nabla_j \chi,
\end{equation}
the electric and magnetic components respectively.
From this equation, one has $\nabla^2 \Phi = 
\nabla^i V_i$, and $\nabla^2 \chi = -\nabla^i \epsilon_i^{\,\,\,j} V_j$.
For a manifold without a boundary, like the full sky, the 
Laplacian may be inverted up to a constant zero mode, 
and the two contributions to $V_i$ are uniquely determined. 
But for a finite patch, 
one can always think of adding charged sources for 
the potentials $\Phi$ and $\chi$ outside of the patch on which
$V_i$ is measured, which alter $\Phi$ and $\chi$ without changing
$V_i$. 
For example one can add to $\Phi$ and $\chi$ pieces with
equal but perpendicular gradients so there is
no net contribution to $V_i$. 

Since full sky observations are unrealistic, so is the hope of
a unique decomposition of the sky polarization into 
electric
and magnetic components. However, this does not
at all mean that the hypothesis of a magnetic signal 
cannot be tested. One possibility is to construct a local measure
of the magnetic signal by differentiating the measured polarization (i.e.\
 $\nabla^i \epsilon_i^{\,\,\,j} V_j$ vanishes if $V_i$ is pure
electric in the analogue example above), but this is problematic 
for noisy, sampled data. A more promising alternative, which avoids
differentiating the data, is to construct line integrals of the
polarization~\cite{Chiueh01, Zaldarriaga01}. For example, in the vector
analogy above, any line integral $\oint\text{d}x^i V_i$ is guaranteed to vanish
if $V_i$ is purely electric.
However, the problem with these line integrals is that there
are an infinite number of them, and they are
not statistically independent. One would therefore prefer
a set of `magnetic' variables to which the 
`electric' component does not contribute,
but which are finite in number and
statistically independent, for a rotationally symmetric statistical
ensemble.
Since polarization from a primordial scale invariant spectrum of 
gravitational waves 
is predominantly generated on scales of a degree or so 
(the angle subtended by the horizon at last scattering),
we would expect to be able to
characterize the cosmic magnetic signal by a set of 
statistically independent variables roughly equal
in number to the size of the patch in square degrees. However the signal
within a degree or so of the boundary cannot be unambiguously
characterized as magnetic, and hence one loses a number of variables
proportional to the length of the boundary. The amount of information
about the magnetic signal therefore increases as the patch area minus
the area of this boundary layer.
In this paper we shall find the set of 
observable `magnetic' variables explicitly for circular sky patches:
the method may be generalized to non-circular patches if necessary.

As mentioned above, the electric component of the 
polarization (due primarily to scalar
perturbations) is expected to be much larger
than the magnetic signal. 
Therefore to detect the latter it may be useful to construct observables which
suffer no electric contamination. 
We show how to construct such
variables, and use them to estimate what magnitude of 
magnetic signal the planned
Planck satellite\footnote{\link{http://astro.estec.esa.nl/Planck}}
might be able to detect. We also discuss the optimal
survey size for future experiments aimed at detecting tensor modes via
magnetic polarization, including the effects of `magnetic noise' due to weak
lensing of the dominant electric polarization~\cite{Guzik00,Hu01}.
Even for observations that do not
expect to detect the magnetic signal the 
magnetic-only observables are likely to be very useful in checking
consistency of any residual polarization with noise
or indeed in identifying foreground contamination. They may also be
useful for studying the small scale weak lensing signal~\cite{Benabed01}.

To construct variables that depend only on the electric or magnetic
polarization we integrate the polarization field over the observed patch
with carefully chosen spin-weight 2 window functions. We present a
harmonic-based approach for constructing these window functions which is
exact in the limit of azimuthally-symmetric patches. The method is
expected still to perform well for arbitrary shaped patches of the sky, but
the separation will no longer be exact in that case. Constructing the
window functions with our harmonic method automatically removes
redundancy due to the finite size of the patch, keeps the information loss
small (except for very small patches), and ensures that for
idealized noise in the polarization map (isotropic and uncorrelated), the
noise on the electric and magnetic variables preserves these properties.
In this respect the construction is analogous to the orthogonalized
harmonics approach used in the analysis of temperature
anisotropies~\cite{Gorski94,Mortlock00}. However in the polarized case there
is no simple interpretation in terms of a set of orthogonalized
harmonics.

In Ref.~\cite{Tegmark00} it was shown how the lossless quadratic
estimator technique can be applied to polarization. There, no attempt was
made to separate the electric and magnetic contribution to the estimators,
so the resulting window functions for the power displayed considerable leakage
between the electric and magnetic modes. The authors of Ref.~\cite{Tegmark00}
showed how the leakage could be reduced, but it is arguably still
too large to allow robust estimation of the magnetic signal in the
presence of an electric signal that is orders of magnitude
larger. We are able to perform a much cleaner separation at the level of the
harmonic components in the map, and as we shall see the
information loss in our approach is quite small for full sky surveys
with a galactic cut.

The electric-magnetic decomposition of the polarization field is exactly
analogous to the corresponding decomposition of projected galaxy ellipticities induced by weak lensing~\cite{Kaiser92,Stebbins96}.
Ref.~\cite{Crittenden00} shows how to  construct local real-space
correlation functions for measuring the magnetic component. These are useful for distinguishing
the purely electric signal due to gravitational lensing from intrinsic
correlations in galaxy alignments, and the method has the advantage of
working for arbitrarily shaped
regions of sky.  However the method assumed a flat sky
approximation, and includes only the two-point information.
For polarization observations the
sky curvature will be important and we aim to extract a set of
statistically independent observables that contain as much of the
magnetic information as possible. This may also prove useful for weak
lensing studies.

The paper is arranged as follows. In Sec.~\ref{sec:ebpol} we present the
spin-weight 2 window technique for separating electric and magnetic
polarization on the sphere, generalizing results in
Refs.~\cite{Chiueh01, Zaldarriaga01}. Section~\ref{sec:sep} describes our
harmonic-based technique for constructing window functions with the properties
required to ensure separation of the electric and magnetic modes while
keeping information loss small. Classical techniques for
testing the hypothesis that there is no magnetic signal are discussed in
Sec.~\ref{sec:hyp}, and estimates of the detection limits with the
Planck satellite and future experiments are also given. Lossless methods for
estimation
of the polarization power spectra are contrasted with methods using the
separated variables in Sec.~\ref{LOSSLESS}. In a series of appendices we
outline our conventions for spin weight functions and their spherical
harmonics. In addition we present a number of the standard integral theorems
on 2-dimensional manifolds in convenient spin weight form, and present
recursive methods for the fast computation of the spin weight spherical
harmonics and their inner products over azimuthally symmetric patches
of the sphere. A further appendix discusses the statistics of
detecting weak signals from tensor modes.

\section{Electric and Magnetic polarization}
\label{sec:ebpol}

The observable polarization field is described in terms of the two
Stokes' parameters $Q$ and $U$ with respect to a particular choice of
axes about each direction on the sky. In this paper we take these axes
to form a right-handed set with the incoming radiation direction (following
Ref.~\cite{Kamionkowski97}). The real Stokes' parameters are
conveniently combined into a single complex field that represents the
observed polarization
\begin{equation}
P = Q + iU.
\end{equation}
The values of Stokes' parameters depend on the choice of axes; since
$Q$ is the difference of the intensity in two orthogonal directions it
changes sign under a rotation of $90^\circ$. The $Q$ field is related
to the $U$ field by a rotation of $45^\circ$. More generally under
a right-handed rotation of the axes by an angle $\alpha$ about the
incoming direction the complex polarization
transforms as $P\rightarrow e^{-2i\alpha} P$ and is therefore 
described as having spin minus two (see Appendix~\ref{App:Eth} for our
conventions for spin weight functions).
The analysis of polarized data is therefore rather
more complicated than for the temperature which does not depend on
making a choice of axes in each direction on the sky.

As described in Appendix~\ref{App:Eth}, one can define spin raising and
lowering operators that can be used to relate quantities of different
spin~\cite{Newman66,Goldberg67}. The
spin raising operator is denoted $\edth$ and the lowering operator
$\beth$. Since the polarization has spin-weight -2 it can be written as
the action of two spin lowering operators on a spin zero complex number
\begin{equation}
P = \beth\beth (P_E + iP_B).
\end{equation}
The underlying real scalar (spin-zero) fields $P_E$ and $P_B$ describe
electric and magnetic polarization respectively~\cite{Newman66}. They
are clearly non-local functions of the Stokes' parameters. One
can define a spin zero quantity which is local in the polarization by acting
with two spin raising operators. Using
some results from Appendix~\ref{App:Eth} one obtains
\begin{equation}
\edth\edth P = (\grad^2+2)\grad^2 (P_E + iP_B)
\end{equation}
where  $\grad$ is the covariant derivative on the sphere.
The real and imaginary parts of this equation can therefore be used to
determine the electric and magnetic parts of the polarization. 
Performing a surface integral we define
\begin{equation}
I'_{-2,W} \equiv \int_S \text{d}S \,W^\ast \,\edth\edth P, \qquad I'_{2,W}
\equiv \int_S \text{d}S\, W^\ast\, \beth\beth P^\ast,
\end{equation}
where $W$ is a complex window function defined over some patch $S$
of the observed portion of the sky. It follows that
\begin{equation}
E'_W \equiv \half (I'_{2,W} + I'_{-2,W}), \qquad B'_W
\equiv -i\half(I'_{2,W} - I'_{-2,W})
\end{equation}
provide a measure of the electric and magnetic
signals. Note that $E'_{W^\ast}=E_W^{\prime\ast}$ with an equivalent result for
$B'_W$.
Using the integral theorem~\eqref{inttheorem2} in Appendix~\ref{App:Eth}
we can write
\begin{eqnarray}
I'_{2,W} &=& \int_S \text{d}S \,P^\ast {}_2 W^\ast + \oint_{\partial S}
\dbeth\left( W^\ast \beth P^\ast - P^\ast \beth W^\ast\right),
\label{eq:i2wp}
\\
I'_{-2,W} &=& \int_S \text{d}S \,P {}_{-2}W ^\ast + \oint_{\partial S}
\deth \left(W^\ast\edth P -P\, \edth W^\ast\right),
\label{eq:im2wp}
\end{eqnarray}
where ${}_2W\equiv\edth\edth W$ is now a spin 2 window function,
${}_{-2}W\equiv \beth\beth W$ is a spin $-2$ window function, and
$\deth = \dbeth^\ast$ is the spin 1 element of length around the boundary
$\partial S$ of $S$. Clearly we do not wish to take derivatives of noisy
observed data and hence it is usually useful to choose the window function to
eliminate the derivative terms on the boundary.

For CMB polarimetry we are interested in the polarization defined on the
spherical sky. The surface integrals vanish if we choose $W$ such that
$\edth\edth W=\beth\beth W=0$, which will be true
if $W$ is a linear combination of the spherical harmonics 
with $l=0$ or 1,
since these possess no spin 2 component. 
If we then set $W=0$ on the boundary, so as to eliminate the
derivatives
of the polarization, we are forced to consider circular patches $S$,
in which case a combination of the two $m=0$ harmonics works. This
implies that the electric and magnetic signals can be probed by performing
line integrals around circles, as emphasized
in Refs.~\cite{Chiueh01,Zaldarriaga01}. These line integrals can be performed
around any circle that is enclosed in the observed region of the sky,
and it is unclear how to obtain a complete
set of statistically independent observables in order to extract all
of the available information. Also for current experiments, performing
one-dimensional line integrals on pixelized maps is unlikely to be a good
way to extract information robustly.

In this paper, we suggest choosing the window functions so that the line
integrals around $\partial S$ that appear in the construction of $E'_W$ and
$B'_W$ contain no contribution from the magnetic and electric polarization
respectively. In the absence of special symmetries (see below for exceptions
that arise in the case of circular patches) this requires that
$W$, $\edth W$, and $\beth W$ all
vanish on the boundary. These conditions are equivalent to demanding that the
window function and its normal derivative vanish on $\partial S$. With
such a choice of window we can measure the
electric and magnetic signals using only the surface integrals
\begin{equation}
I_{\pm2,W} \equiv E_W \pm iB_W \equiv \int_S \text{d} S
\,\, {}_{\pm 2}W^\ast (Q \mp i U).
\label{eq:ewbw}
\end{equation}
Since the window functions are scalar functions on the sphere we can
expand them in spherical harmonics,
\begin{equation}
W = \sum_{l \geq 2} \sum_{|m| \leq l} \sqrt{\frac{(l-2)!}{(l+2)!}}
W_{lm} Y_{lm}.
\label{eq:hwindow}
\end{equation}
(The square root factor is included for later convenience.) We need not
include $l=0$ and 1 spherical harmonics since they do not contribute to the
spin-weight $\pm 2$ window functions, and the boundary integral terms
automatically separate for these multipoles. In practice, we are only
interested in probing scales to some particular $\lmax$ (e.g.\ the
magnetic signal from tensor modes has maximal power for $l\approx 100$ and
decreases rapidly with $l$), so the sum in Eq.~(\ref{eq:hwindow}) can be
truncated at some finite $\lmax$.

We shall focus on the case where the observed sky patch is azimuthally
symmetric in which case the construction of exact window functions becomes
particularly simple. The harmonic-based method we describe in
Sec.~\ref{sec:sep} provides a practical solution to constructing a
non-redundant set of window functions that separate the electric
and magnetic modes exactly. In addition, for the special case of isotropic,
uncorrelated noise on the observed polarization, these simple properties
are preserved in the variables $E_W$ and $B_W$. For observations over
non-azimuthally symmetric patches our method can, of course, be used over the
largest inscribed circular patch, but in this case there is inevitable
information loss since we use only a subset of the observed data. However,
we expect that the method presented in Sec.~\ref{sec:sep} could also be applied
directly to the full observed region to construct window functions that
achieve approximate separation of electric and magnetic polarization.

Consider the case of an azimuthally-symmetric patch so the boundary $\partial
S$ consists of one or two small circles.
For each azimuthal dependence on $m$ we can construct combinations
\begin{equation}
W_m=\sum_{l=\text{max}(2,|m|)}^{\lmax} \sqrt{\frac{(l-2)!}{(l+2)!}}W_{lm}
Y_{lm} \label{eq:hwindowm}
\end{equation}
that satisfy the
necessary boundary conditions. For $m=0$ it is easy to see that
$E_W$ and $B_W$ contain no
contribution from $P_B$ and $P_E$ respectively for any choice of the
$W_{l0}$ [i.e.\ the boundary integrals that distinguish $E_W$ ($B_W$) from
$E'_W$ ($B'_W$) vanish if the polarization is pure magnetic (electric)].
It follows that for $m=0$ there are $\lmax-1$ linearly independent window
functions that satisfy
the boundary conditions. For $|m|=1$ it will be shown in the next section
that there is only one independent linear constraint per boundary circle,
so there are $\lmax-2$ possible window functions ($\lmax-3$ for a boundary
composed of two circles). For $|m| \geq 2$ there are two linear constraints
per boundary circle which can be taken to be the vanishing of $W_m$ and
its normal derivative. In this case there are $\lmax-|m|-1$ ($\lmax-|m|-3$)
window functions for boundaries consisting of one (two) small circles.

Since we are only considering a fraction of the sky not all of the window
functions counted above may return observables $E_W$ and $B_W$ containing
independent information. This arises because for large $\lmax$, or small
patches, there will generally arise non-zero window coefficients $W_{lm}$
that produce spin 2 window functions that are poorly supported over the
patch. (See e.g.\ Ref.~\cite{Mortlock00} for a discussion of the equivalent
problem in the case of scalar functions.) The redundancy in the set
of acceptable window functions can be removed by expanding the
spin 2 window functions in a smaller set of functions which are (almost)
complete for band-limited signals over the region $S$. The construction
of such a set by singular value methods
(e.g.\ Refs.~\cite{MatrixOps,Mortlock00})
forms the starting point of the method we present in the Sec.~\ref{sec:sep}.

\subsection{Harmonic expansion}

We construct window functions in harmonic space, so as a useful preliminary we
consider the harmonic expansion of spin-weight 2 fields over the full
sphere~\cite{Kamionkowski97,Zaldarriaga97}.
The polarization $P\equiv Q + iU$ is spin $-2$ and can be expanded
over the whole sky in terms of the spin two harmonics
(see Appendix~\ref{App:harmonics} for our conventions and some useful results)
\begin{equation}
Q\pm i U = \sum_{lm} a_{\mp 2,lm}\,{}_{\mp 2}Y_{lm} =  \sum_{lm}
(E_{lm} \mp i B_{lm}){}_{\mp 2}Y_{lm}.
\end{equation}
Reality of $Q$ and $U$ requires $a_{-2,lm}^\ast = (-1)^m a_{2,l(-m)}$, so that
$E_{lm}^\ast = (-1)^m E_{l(-m)}$ with an equivalent result for $B_{lm}$.
Under parity transformations $E_{lm} \rightarrow (-1)^l E_{lm}$ but
$B_{lm} \rightarrow (-1)^{l+1} B_{lm}$, since
${}_sY_{lm}(\pi-\theta, \phi+\pi) = (-1)^l {}_{-s}Y_{lm}(\theta,\phi)$.
From the orthogonality of the spherical harmonics over the full sphere
it follows that
\begin{eqnarray}
E_{lm} &=& \half( a_{2,lm} + a_{-2,lm}) = \half\int_{4\pi} \text{d}S\,
P {}_{-2}Y_{lm}^\ast+\half \int_{4\pi} \text{d}S\, P^\ast
{}_{2}Y_{lm}^\ast , \label{eq:elm}\\
i B_{lm} &=& \half( a_{2,lm} - a_{-2,lm}) = - \half \int_{4\pi} \text{d}S\,
P{}_{-2}Y_{lm}^\ast + \half\int_{4\pi} \text{d}S\,P^\ast
{}_2Y_{lm}^\ast . \label{eq:blm}
\end{eqnarray}

In a rotationally-invariant ensemble, the expectation values of the harmonic
coefficients define the electric and magnetic polarization power spectra:
\begin{equation}
\la E_{l'm'}^\ast E_{lm} \ra = \delta_{l'l}\delta_{m'm} C_l^{EE},
\qquad \la B_{l'm'}^\ast B_{lm} \ra = \delta_{l'l}\delta_{m'm} C_l^{BB}.
\end{equation}
If the ensemble is parity-symmetric the cross term is zero,
$\la E_{l'm'}^\ast B_{lm} \ra=0$.

\begin{figure}
\begin{center}
\epsfig{figure=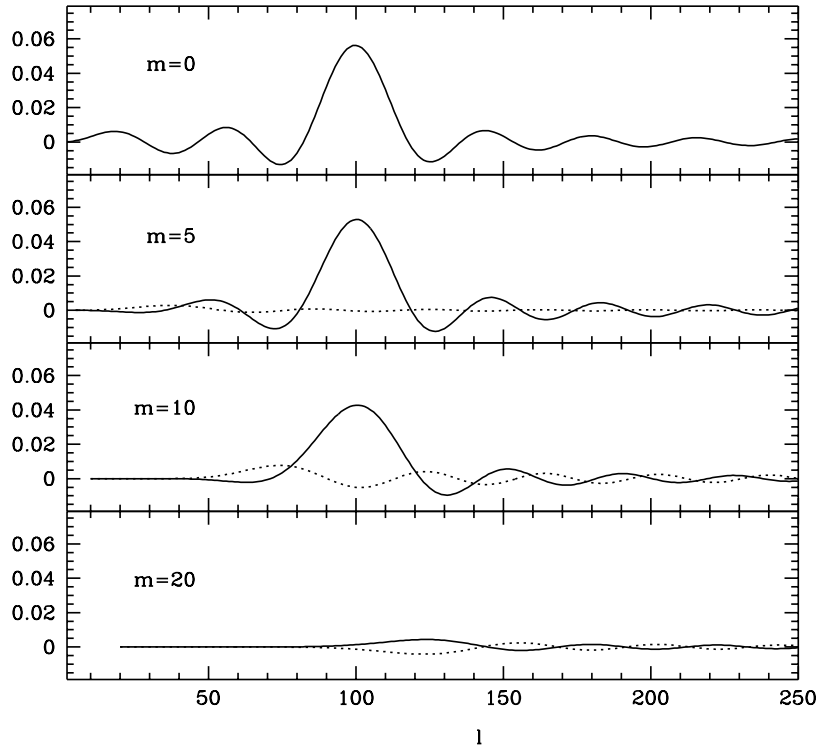,angle=0,width=12.5cm} 
\caption{The window functions $W_{+(l'm)(lm)}$ (solid lines) and
$W_{-(l'm)(lm)}$ (dashed lines) for $l'=100$ and various $m$ for an
azimuthally symmetric patch with $\theta < 10^\circ$. The
dashed lines show the $E_{lm}$ contamination of $\Bt_{l'm}$ as a
function of $l$. For $m=0$ there is no contamination, and as $m$
increases the functions decrease in amplitude as the corresponding
harmonics become more localized outside of the patch. \label{windows}}
\end{center}
\end{figure}

The form of the harmonic expansion~(\ref{eq:hwindow}) of the window function
ensures that the spin-weight $\pm 2$ windows are
\begin{equation}
{}_{\pm 2}W = \sum_{lm} W_{lm} {}_{\pm 2}Y_{lm},
\end{equation}
where the sum is over $l\geq 2$ and $|m| \leq l$. Evaluating the surface
integrals in Eq.~(\ref{eq:ewbw}) we find
\begin{equation}
E_W  = \sum_{lm} W_{lm}^\ast \tilde{E}_{lm}, \qquad B_W = \sum_{lm}
W_{lm}^\ast \tilde{B}_{lm}, 
\label{eq:ewbw_harm}
\end{equation}
where the pseudo-harmonics are obtained by restricting the integrals
in Eqs.~(\ref{eq:elm}) and (\ref{eq:blm}) to the region $S$:
\begin{eqnarray}
\tilde{E}_{lm} &=& \frac{1}{2}\sum_{l'm'}\int_S\text{d}S\, \left[ (E_{l'm'} - i
B_{l'm'}){}_{-2}Y_{l'm'}{}_{-2}Y_{lm}^\ast + (E_{l'm'} + i B_{l'm'})
{}_2Y_{l'm'}{}_2Y_{lm}^\ast \right] , \label{nosepeqe}
\\
\tilde{B}_{lm} &=& \frac{i}{2}\sum_{l'm'}\int_S \text{d}S\,
\left[ (E_{l'm'} - i
B_{l'm'}){}_{-2}Y_{l'm'}{}_{-2}Y_{lm}^\ast - (E_{l'm'} + i B_{l'm'})
{}_2Y_{l'm'}{}_2Y_{lm}^\ast  \right]. 
\label{nosepeqb}
\end{eqnarray}
Defining Hermitian coupling matrices
\begin{equation}
W_{\pm(lm)(lm)'} \equiv \half({}_2W_{(lm)(lm)'} \pm {}_{-2}W_{(lm)(lm)'}),
\end{equation}
where
\begin{equation}
{}_sW_{(lm)(lm)'} \equiv \int_S \text{d}S \, {}_s Y_{lm}^\ast {}_s Y_{l'm'},  
\end{equation}
we can write
\begin{eqnarray}
\tilde{E}_{lm} &=& \sum_{l'm'} \left(W_{+(lm)(lm)'} E_{l'm'} + i W_{-(lm)(lm)'}
B_{l'm'}\right), \\
\tilde{B}_{lm} &=& \sum_{l'm'} \left(W_{+(lm)(lm)'} B_{l'm'} - i W_{-(lm)(lm)'}
E_{l'm'}\right). 
\end{eqnarray}
In the limit $\lmax \rightarrow \infty$, ${}_{\pm 2} W_{(lm)(lm)'}$ become
projection operators as a consequence of the completeness of the spin-weight
harmonics.
The matrix $W_{-(lm)(lm)'}$ controls the contamination of $E_W$ and $B_W$
with magnetic and electric polarization respectively. Our aim is to construct
window functions $W_{lm}$ that remove this contamination for all $E_{lm}$ and
$B_{lm}$. Some elements of the matrices $W_{\pm(lm)(lm)'}$
are shown in Fig.~\ref{windows}.

For azimuthally-symmetric patches the coupling matrices are block
diagonal ($W_{\pm (lm)(lm)'} \propto \delta_{mm'}$), and so window functions
can be constructed for each $m$ separately [see Eq.~(\ref{eq:hwindowm})].
For $m=0$ we have
\begin{equation}
\int_S \text{d}S {}_2Y_{l'0}\,{}_2Y_{l0}^\ast =  \int_S \text{d}S
{}_{-2}Y_{l'0}\,{}_{-2}Y_{l0}^\ast,
\end{equation}
so $W_{-(l0)(l'0)}=0$ and we have clean separation for any
azimuthally-symmetric window function. The set of azimuthally symmetric
window functions gives $\lmax -1$ separated  variables that contain
the same information as would be obtained by computing
line integrals around all those circles concentric with the boundary of the
azimuthal patch. For general $m$ there is leakage
of $E_{lm}$ into $\tilde{B}_{lm}$; for parity-symmetric cuts there is
only leakage between modes with different parity (i.e.\ for even $l$
the pseudo-harmonics $\tilde{B}_{lm}$ depend on $E_{l'm}$ only for odd $l'$).

We showed in the previous section that, for a general window function,
the contamination of e.g.\ $E_W$ by the magnetic polarization is due entirely
to boundary terms. This implies that $W_{-(lm)(lm)'}$ can always be written as
a line integral around the boundary of $S$. (We show in Appendix~\ref{App:Ints}
that the matrices ${}_s Y_{(lm)(lm)'}$ can be transformed into line integrals
for $l \neq l'$. However $W_{-(lm)(lm)'}$ can be written as a line integral
for all $l$ and $l'$.) Making use of
Eq.~(\ref{eq:appcdint}), it is straightforward to show that
\begin{eqnarray}
W_{-(lm)(lm)'} &=& \frac{1}{2}\sqrt{\frac{(l-2)!}{(l+2)!}} \Bigl(
\oint_{\partial S} \dbeth\, \left[\sqrt{l(l+1)} {}_1 Y_{lm}^\ast
{}_2Y_{l'm'} + \sqrt{(l'-1)(l'+2)} Y_{lm}^\ast {}_1 Y_{l'm'} \right]\nonumber\\
&&\mbox{}+ \oint_{\partial S} \deth\, \left[\sqrt{l(l+1)} {}_{-1}Y_{lm}^\ast
{}_{-2}Y_{l'm'} + \sqrt{(l'-1)(l'+2)} Y_{lm}^\ast {}_{-1}Y_{l'm'} \right]
\Bigr).
\label{eq:wminusint}
\end{eqnarray}
This can be put in manifestly Hermitian form using the recursion
relation derived from the action of $(\edth -\beth)$ on ${}_sY_{lm}$
\begin{equation}
\frac{m+s\cos\theta}{\sin\theta} {}_sY_{lm} = \half \sqrt{(l-s)(l+s+1)}
{}_{s+1}Y_{lm} + \half \sqrt{(l+s)(l-s+1)} {}_{s-1}Y_{lm}.
\end{equation}
For a circular boundary at constant latitude $\theta=\Theta$
(i.e.\ the boundary of an azimuthal patch), we find
\begin{equation}
W_{-(lm)(l'm)} = - 4 m\pi [u_l(m) u_{l'}^\ast(m) + v_l(m) v_{l'}^\ast(m)],
\label{eq:wminusdcmp}
\end{equation}
where the vectors
\begin{eqnarray}
u_l(m) &=& \sqrt{\frac{(l-2)!}{(l+2)!}}\left[
\partial_\theta Y_{lm}(\Theta,\Phi) - \cot\Theta Y_{lm}(\Theta,\Phi)
\right], \\
v_l(m) &=& \sqrt{\frac{(l-2)!}{(l+2)!}}\frac{\sqrt{(m^2-1)}}{\sin\Theta} Y_{lm}(\Theta,\Phi)
\end{eqnarray}
for $l\geq 2$ and some arbitrary $\Phi$.
[Note that $u_l(m)$ and $v_l(m)$ will not generally
be orthogonal so Eq.~(\ref{eq:wminusdcmp}) is not the spectral decomposition
of $W_{-(lm)(l'm)}$.]
Any window $W_{lm}$ whose inner products with
$u_l(m)$ and $v_l(m)$ both vanish, i.e.\
\begin{equation}
\sum_l W_{lm}^\ast u_l(m) = \sum_l W_{lm}^\ast v_l(m)  = 0, 
\end{equation}
will achieve clean separation of electric and magnetic polarization. For
$|m| > 1$ such window functions and their normal derivative necessarily
vanish on the boundary. As noted earlier, for $|m|=1$ there is actually only
one constraint to be satisfied which now follows from the fact that
$v_l(\pm1)=0$.

\section{Electric/Magnetic separation on the cut sky}
\label{sec:sep}

In this section we give a practical method for constructing a non-redundant set
of window functions $\{ W_I \}$ where $I$ labels the particular window,
that achieve exact separation for azimuthal patches. The corresponding
(cleanly separated) electric and magnetic observables will
be denoted $E_{W_I}$ and $B_{W_I}$. We will make use of a notation where
vectors are denoted by bold Roman font, e.g.\ $\vBw$ has components
$B_{W_I}$, and $\vB$ has components $B_{lm}$, and matrices are denoted by
bold italic font, e.g.\ $\mW_\pm$ have components $W_{\pm(lm)(lm)'}$.
We present the method in a form that is applicable (though no longer exact)
to arbitrary shaped regions $S$; for azimuthal patches the method is exact.
For the azimuthal case all matrices are block
diagonal and the window functions can be constructed for each $m$ separately.

In matrix form, Eq.~(\ref{eq:ewbw_harm}) is
\begin{equation}
\vEw = \mW^\ast \vEt, \qquad \vBw = \mW^\ast \vBt,
\label{eq:meb}
\end{equation}
where $\mW = W_{I(lm)}$ is the matrix whose $I$th row contains the harmonic
coefficients of the $I$th window function, and recall
\begin{equation}
\vEt = \mW_+ \vE + i \mW_- \vB, \qquad \vBt =  \mW_+ \vB  - i
\mW_- \vE.
\end{equation}
For an azimuthally-symmetric sky patch the block-diagonal matrices
${}_{\pm 2} \mW$ (with components ${}_{\pm 2}W_{(lm)(l'm)}\delta_{mm'}$)
from which $\mW_\pm$ are constructed can be
computed very quickly using the recursion relations given in
Appendix~\ref{App:Ints}. Alternatively, $\mW_-$ can be computed
directly from Eq.~(\ref{eq:wminusdcmp}). In the limit of full sky coverage
$\mW_- \rightarrow 0$ and $\mW_+ \rightarrow \mI$. We know that for
$|m|\ge 2$, the range of $m$th submatrix of $\mW_-$ is two-dimensional [spanned
by $u_l(m)$ and $v_l(m)$], so that all but two of the
eigenvalue of the submatrix are zero. Equivalently, all but two linear
combinations of the $\tilde{B}_{lm}$ are independent of $E_{l'm}$. 
The $|m|=1$ submatrices of $\mW_-$ have only one non-zero eigenvalue; the
associated eigenvectors are $u_l(\pm 1)$. The $m=0$ submatrix is identically
zero. The essence of our method for constructing the window functions
is to choose $\mW$ to project out of the range of $\mW_-$.

\begin{figure}
\begin{center}
\psfig{figure=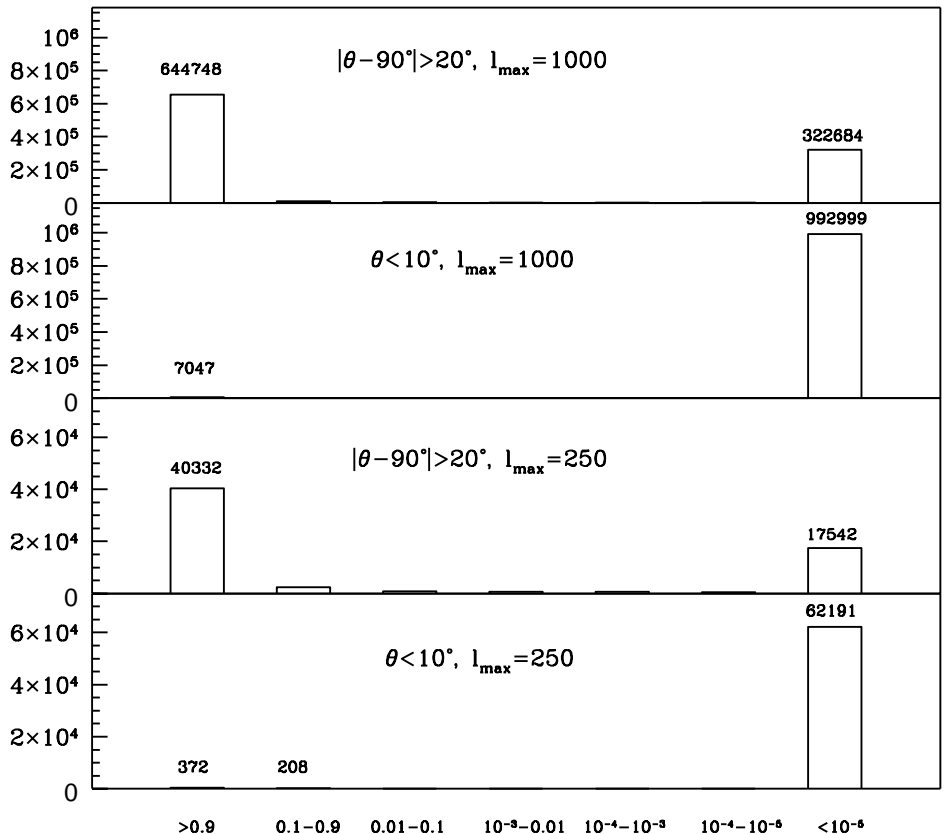,angle=0,width=14.5cm} 
\caption{The distribution of eigenvalues of $\mW_+$ for two
azimuthally-symmetric sky cuts with $\lmax=
\{250,1000\}$. The distribution is approximately bimodal, and the
fraction of the eigenvalues corresponding to well-determined modes
(eigenvalue significantly non-zero)
is given by the fraction of the sky area in the patch in the limit
$\lmax\rightarrow\infty$. \label{evals}}
\end{center}
\end{figure}

We first diagonalize $\mW_+ =\mU_+ \mD_+ \mU^\dagger_+$ by performing
a singular value decomposition~\cite{MatrixOps}. Here,
$\mD_+$ is a positive (semi-)definite diagonal matrix whose elements are
the eigenvalues of $\mW_+$. The columns of the unitary matrix $\mU_+$ are the
normalized eigenvectors of $\mW_+$. The singular value decomposition
allows us to identify the linear combinations $\mU^\dagger_+ \vB$ that are
poorly determined by $\vBt$ --- those
corresponding to the small diagonal elements of $\mD_+$. The eigenvectors
with very small eigenvalues correspond to polarization patterns that
essentially have no support inside the observed patch of the sky, and
would lead to a set of redundant window functions if not removed from the
analysis. The
distribution of eigenvalues of $\mW_+$ is approximately bimodal as
illustrated in Fig.~\ref{evals}, and the exact definition of `small' is not
critical when considering the range of the matrix. This bimodality arises
because ${}_{\pm 2}\mW$ are approximately projection operators for large
$\lmax$, and the fact that the range of $\mW_-$ is a rather small subspace.
To remove redundant degrees of freedom from the spin 2 window functions,
we define an operator $\mUt_+$ which projects onto the eigenvectors of
$\mW_+$ whose eigenvalues are close to one. This amounts to removing
the appropriate columns of $\mU_+$. Since $\mU_+$ is orthogonal,
$\mUt_+$ is column orthogonal and hence $\mUt^\dagger_+ \mUt_+ = \mI$
(but $\mUt_+ \mUt_+^\dagger \ne \mI$). The matrix $\mDt_+$ is the
corresponding smaller square diagonal matrix, and we have
\begin{equation}
\mUt^\dagger_+ \vBt \approx \mDt_+ \mUt^\dagger_+ \vB - i\mUt^\dagger_+
\mW_- \vE.
\end{equation}
We now multiply by $\mDt_+^{-1/2}$, defined by $[\mDt_+^{-1/2}]_{ij}
\equiv \delta_{ij}[ \mDt_+]_{ii}^{-1/2}$, to give
\begin{equation}
\mDt_+^{-1/2}\mUt^\dagger_+ \vBt \approx \mDt_+^{1/2} \mUt^\dagger_+
\vB - i\mDt_+^{-1/2}\mUt^\dagger_+ \mW_- \vE.
\end{equation}
Including the factor $\mDt_+^{-1/2}\mUt^\dagger_+$ in the window functions
$\mW$ is equivalent to constructing the spin 2 window functions from
a reduced basis that is orthonormal and (almost) complete over the region
$S$.

The remaining step is to project out the unwanted boundary terms that
contaminate $\vBt$ with $\vE$. For the case of an azimuthal patch, the
ranges of the submatrices of $\mW_-$ are all two dimensional (or lower).
It follows that the same is true of $\mDt_+^{-1/2}\mUt^\dagger_+ \mW_-$,
so we can remove the unwanted boundary term by ensuring that $\mW$ also
includes a factor that projects out of the range of
$\mDt_+^{-1/2}\mUt^\dagger_+ \mW_-$. In practice, we perform this projection
by constructing its singular value decomposition, which for a non-square
matrix takes the form $\mDt_+^{-1/2}\mUt^\dagger_+ \mW_- =
\mU\mD\mV^\dagger$. Here, $\mD$ is a diagonal matrix with the same
dimension as $\mDt_+$, $\mU$ is a unitary matrix of the same dimension,
and $\mV$ is a column orthogonal rectangular matrix. There are at most
two non-zero singular values (elements of the diagonal matrix $\mD$) per
$m$, and the corresponding left singular vectors (columns of $\mU$)
form an orthonormal basis for the range of $\mDt_+^{-1/2}\mUt^\dagger_+ \mW_-$.
We can project out of this range by defining $\mUt$ as the matrix obtained by
removing the columns of $\mU$ where the corresponding singular value
is non-zero. Thus, choosing the window functions
\begin{equation}
\mW^\ast = \mUt^\dagger \mDt_+^{-1/2}\mUt^\dagger_+,
\end{equation}
we guarantee separation of the electric and magnetic polarization for
azimuthally-symmetric patches. Our separated polarization observables
become
\begin{eqnarray}
\vBw &=& \mW^\ast \vBt \equiv \mUt^\dagger \mDt_+^{-1/2}\mUt^\dagger_+ \vBt
\approx \mUt^\dagger \mDt_+ ^{1/2}\mUt^\dagger_+ \vB ,\label{eq:bvar}\\
\vEw &=& \mW^\ast \vEt \equiv \mUt^\dagger \mDt_+^{-1/2}\mUt^\dagger_+ \vEt
\approx \mUt^\dagger \mDt_+ ^{1/2}\mUt^\dagger_+ \vE. \label{eq:evar}
\end{eqnarray}
For azimuthal patches the separation is exact; the approximation sign
arises only from our use of $\mW_+ \approx \mUt_+ \mDt_+ \mUt_+^\dagger$
in simplifying the matrix that premultiplies $\vE$ and $\vB$ in
Eqs.~(\ref{eq:bvar}) and (\ref{eq:evar}).

For observations over non-azimuthal patches, one can either apply the
exact separation over the largest inscribed azimuthal region or attempt
to apply the method outlined above to the entire patch. In the latter case
the structure of $\mW_-$ is less clear, but we can still expect
a significant number of its eigenvalues to be very small; the associated
eigenvectors correspond to window functions that satisfy
$W \approx 0 \approx \edth W$ on the boundary. The number of
independent window functions that achieve separation can be estimated as
the number of pixels of linear size $\sim \pi/\lmax$ contained in the patch
(roughly the number of modes that survive the diagonalization of
$\mW_+$) minus twice the number of pixels on the boundary (roughly the
number of constraints in setting $W$ and $\edth W$ to zero on the boundary).
Such window functions will only give an approximate separation of electric and
magnetic polarization. In practice,
the accuracy of the separation, and the number of independent window
functions constructed, will depend on the choice of threshold for retaining
the singular values of $\mDt_+^{-1/2}\mUt^\dagger_+ \mW_-$.

If we are only interested in constructing variables that depend on
the magnetic polarization, the maximum multipole $\lmax$ in the window
functions can be chosen rather small (of the order of a few hundred).
The relation $\vBt = \mW_+ \vB - i \mW_- \vE$ will only hold for square
$\mW_-$ if $\lmax$ is chosen to include all the significant power in the
electric polarization, so for smaller $\lmax$ we cannot assume that
$\mW_-$ is square. However, for azimuthal patches, the range of each
submatrix of $\mW_-$ is still guaranteed to be of dimension two or less,
irrespective of its shape, and so the exact separation can proceed with
$\mW_-$ treated as square. For non-azimuthal patches it would be prudent
to monitor the effect of varying the number of columns in $\mW_-$ on the
range of this matrix.

\subsection{Noise}

It is straightforward to project the errors on the Stokes' parameters
to find the noise in the separated variables $\vEw$ and $\vBw$. In this
section we consider the simple case of maps with idealized noise properties.

We assume that the noise correlation between pixels is negligible, and
that the noise on the Stokes' parameters $\Delta Q$ and $\Delta U$
is un-correlated.
The neglect of noise correlations between pixels amounts to assuming white
noise in the time stream of the measurement. In principle, correlations
between Stokes' parameters can be eliminated with a careful choice of
polarimeter directions in the experiment~\cite{Couchot99,Challinor01}.
With these assumptions, we have
\begin{equation}
\la \Delta Q(\Omega) \Delta U(\Omega') \ra =0,
\end{equation}
and the noise correlation is given by
\begin{equation}
\la \Delta Q(\Omega) \Delta Q(\Omega')\ra = \sigma^2_Q\delta(\Omega-\Omega'),
\qquad \la \Delta U(\Omega) \Delta U(\Omega') \ra
= \sigma^2_U\delta(\Omega-\Omega').
\end{equation}
It follows that the noise $\Delta \vBt$ on the pseudo-multipoles $\vBt$ has
correlations 
\begin{equation}
\la \Delta \vBt \Delta \vBt^\dag \ra = \half \int_S \text{d}S\,
(\sigma^2_Q+\sigma_U^2)\half [{}_{-2} \vY^\ast
({}_{-2}\vY^\ast)^\dag + {}_2 \vY^\ast ({}_2 \vY^\ast)^\dag]
- \half \int_S \text{d}S\,
(\sigma^2_Q - \sigma^2_U)\half[{}_{-2} \vY^\ast
({}_{2}\vY^\ast)^\dag + {}_2 \vY^\ast ({}_{-2} \vY^\ast)^\dag],
\label{eq:noise1}
\end{equation}
with a similar result for the noise $\Delta \vEt$ on $\vEt$. Here, vectors of
spin-weight $s$ functions ${}_s\vY$ have components ${}_sY_{lm}$.
The polarimeter arrangements that give uncorrelated errors between
Stokes' parameters also ensure that
$\sigma^2_U=\sigma^2_Q\equiv \sigma^2_N$ so that the last integral in
Eq.~(\ref{eq:noise1}) is zero.  Here we concentrate on the simple case where $\sigma^2_N$ is isotropic in which case
\begin{equation}
\la \Delta \vBt \Delta \vBt^\dag \ra = \la \Delta \vEt \Delta \vEt^\dag \ra 
= \sigma^2_N \mW_+.
\end{equation}
The covariance of the noise $\Delta \vBw$ on $\vBw$
is therefore given by 
\begin{equation}
\mN \equiv \la \Delta \vBw \Delta \vBw^\dagger \ra =
\mW^\ast \la \Delta \vBt \Delta \vBt^\dag \ra (\mW^\ast)^\dagger \approx
\sigma^2_N \mUt^\dagger  \mDt_+^{-1/2}\mUt^\dagger_+ \mUt_+
\mDt_+\mUt^\dagger_+ \mUt_+ \mDt_+^{-1/2}\mUt = \sigma^2_N\mI,
\end{equation}
and hence the noise is diagonal (and isotropic); similarly $\la \Delta\vEw \Delta\vEw^\dagger
\ra = \sigma^2_N \mI$. What is more, the noise on $\vEw$ and $\vBw$ are
uncorrelated for isotropic noise since 
\begin{equation}
\la \Delta \vEt \Delta \vBt^\dag \ra = i\sigma^2_N \mW_- ,
\end{equation}
and hence
\begin{equation}
\la \Delta\vEw \Delta\vBw^\dag \ra = i\sigma^2_N 
\mUt^\dagger  \mDt_+^{-1/2}\mUt^\dagger_+\mW_- \mUt_+
\mDt_+^{-1/2}\mUt = 0
\end{equation}
as $\mUt^\dagger$ annihilates $\mDt_+^{-1/2}\mUt^\dagger_+\mW_-$.
For isotropic noise our polarization variables therefore have the same
desirable diagonal properties as the scalar diagonalized harmonic
coefficients used in the analysis of the cut-sky CMB
temperature~\cite{Gorski94,Mortlock00}.

In the presence of a symmetric beam, white noise in the time stream of the
experiment projects to white (though generally non-isotropic) noise
on the beam-convolved polarization field. In multipole space, the convolved
fields have electric and magnetic multipoles that are related to the
unconvolved $E_{lm}$ and $B_{lm}$ by spin-weight 2 beam window functions
${}_2 W_l$~\cite{Ng99,Challinor00}. For $\lmax$ of a few hundred, appropriate
for probing magnetic polarization, the beam window functions will be
negligible for experiments with resolution much better than
one degree. For lower resolution experiments, or higher $\lmax$, the effect
of the beam window function should be included in the theoretical (signal)
covariance of the variables $\vEw$ and $\vBw$ (see below).

In general, non-uniform coverage of the sky will lead to variations in
$\sigma^2_N$. In this case it is still possible to define harmonic variables
 $\vBw^{\text{diag}} = \mN^{-1/2} \vBw$ that have isotropic noise. (Here
$\mN^{-1/2}=\mU_N\mD_N^{-1/2}\mU_N^\dag$ where $\mN=\mU_N\mD_N\mU_N^\dag$, $
\mU_N$ is unitary and $\mD_N$ is diagonal.) For azimuthal patches
with general (everywhere finite) noise patterns we can still construct the magnetic-only
variables for each value of $m$ but the noise will now couple variables
with different $m$ (unlike the signal). While presenting no fundamental
obstacles, this $m$-mode coupling does increase the computational
overhead considerably. 

Sky patches of general shape are equivalent to
azimuthal patches with regions of infinite noise, and the general
case of a non-azimuthally symmetric survey region with arbitrary noise can be
treated by re-defining the pseudo-harmonics $\vEt$, $\vBt$ and coupling matrices
$\mW_+$, $\mW_-$ with a factor of $1/\sigma_N^2$ inside the
integral. All the above results then follow with the new definitions,
though computing the matrices and manipulating them may become
computationally challenging. The assumption of isotropic
noise is therefore not fundamental to our analysis, and the following
results could be generalized for more realistic situations.

\begin{figure}[t!]
\begin{center}
\psfig{figure=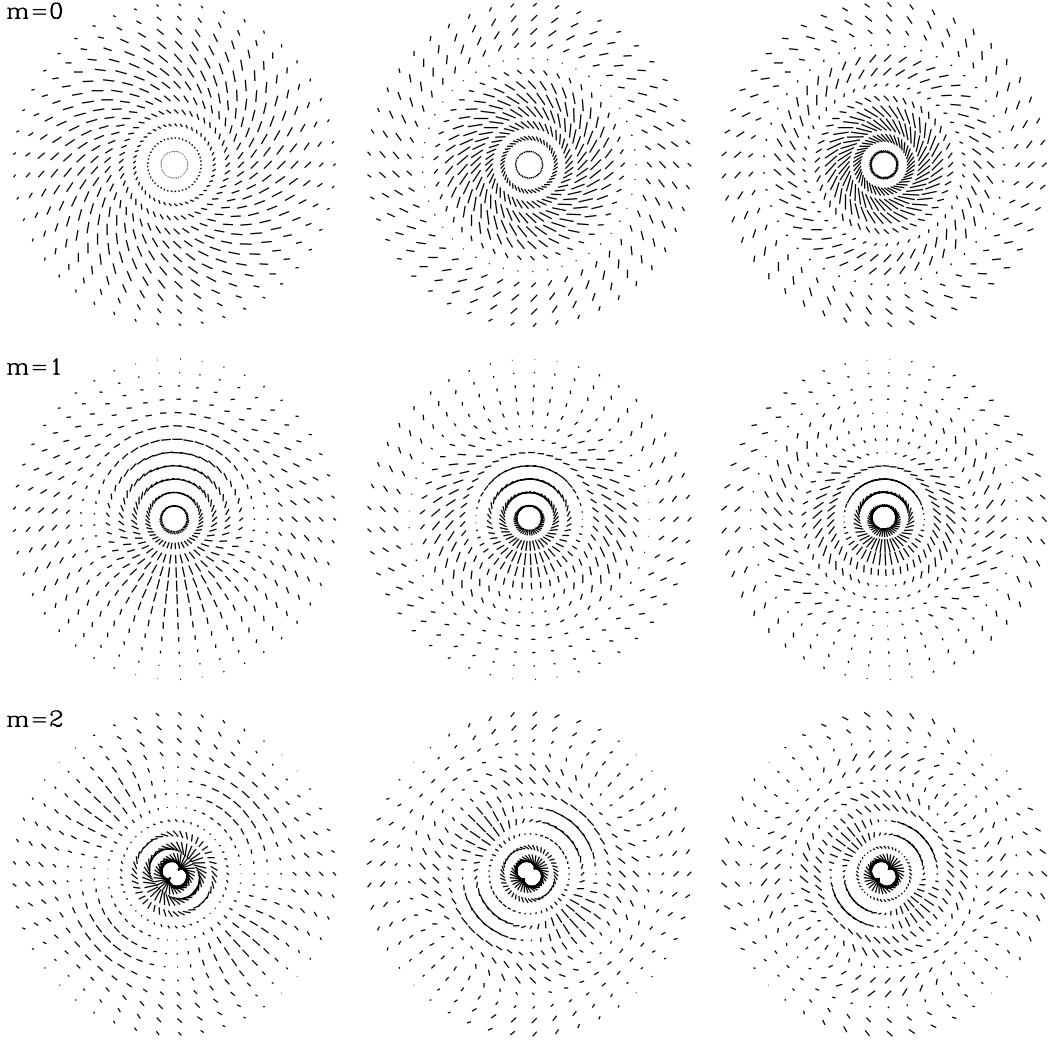,angle=0,width=14.5cm} 
\caption{The real space window functions for an azimuthally symmetric
sky patch with $\theta<10^\circ$. They are evaluated in the frame
where the signal is diagonal, so the leftmost window produces the
largest signal for that $m$. The signal to noise falls off as a function of $m$
as shown in Fig.~\ref{contribs}. For $m>0$ the window functions correspond to
the real part of $\vBwr$; the imaginary part is a rotated version
of the same window. The length of the lines shows the sampling weight
at that point, and the orientation of the lines shows which
polarization orientation gives maximal contribution.
\label{realspace}}
\end{center}
\end{figure}

\subsection{Real space window functions}
The expected magnetic signal correlation matrix is given by
\begin{equation}
\mS \equiv  \la \vBw \vBw^\dag \ra = \mUt^\dagger\mDt^{1/2}_+\mUt^\dagger_+
\mC^{BB} \mUt_+ \mDt_+^{1/2} \mUt ,
\end{equation}
where the diagonal magnetic power spectrum matrix is given by
$[\mC^{BB}]_{(lm)(l'm')}=\delta_{m m'}\delta_{l l'}C_{l}^{BB}$.
Since the noise correlation is proportional to the identity matrix for
isotropic noise we
can perform any rotation $\vBw \rightarrow \vBwr \equiv \mR \vBw$,
where $\mR$ is unitary,
and still have a set of variables with uncorrelated errors. The rotated
variables are derived from window functions $\mR^\ast \mW$.
For a particular theoretical model we can rotate to the frame where the
signal matrix is diagonal. The rotated $\vBwr$ will then be fully
statistically independent. In Fig.~\ref{realspace} we plot the window
functions for the $\vBwr$ which give the largest contributions to the
signal for a typical flat $\Lambda$CDM model with a scale invariant
tensor initial power spectrum and no reionization. The window functions
are plotted as line segments of length $\sqrt{Q_W^2 + U_W^2}$ at angle
$\tan^{-1}(U_W/Q_W)/2$ to the $\theta$ direction where the real 
quantities $Q_W$ and $U_W$ are defined in terms of the real part $\Re \RW$
of the (rotated) scalar window function $\RW$ as
\begin{equation}
Q_W + i U_W \equiv -i \beth\beth \Re \RW.
\end{equation}
This definition ensures that
\begin{equation}
\Re \vBwr = \int_S \text{d}S\left( Q_W Q + U_W U\right).
\end{equation}
For the imaginary part of $\vBwr$, $Q_W$ and $U_W$ should be defined
as $Q_W + i U_W \equiv -i \beth\beth \Im \RW^{\ast}$.
For the
case of azimuthal patches, as considered in Fig.~\ref{realspace} where
the windows are constructed for each $m$, the imaginary part
would produce a plot that is rigidly rotated by $-\pi/(2m)$ ($m\neq 0$)
about the centre.
Plotting the window functions in this form is useful since the length
of the line segment gives the sampling weight assigned to that point, and the
orientation gives the direction of the linear polarization that contributes
at each point.  We could repeat the exercise for the $\vEw$ in which case
for the real part we would define $Q_W + i U_W \equiv \beth\beth \Re \RW$.

\begin{figure}[t!]
\begin{center}
\psfig{figure=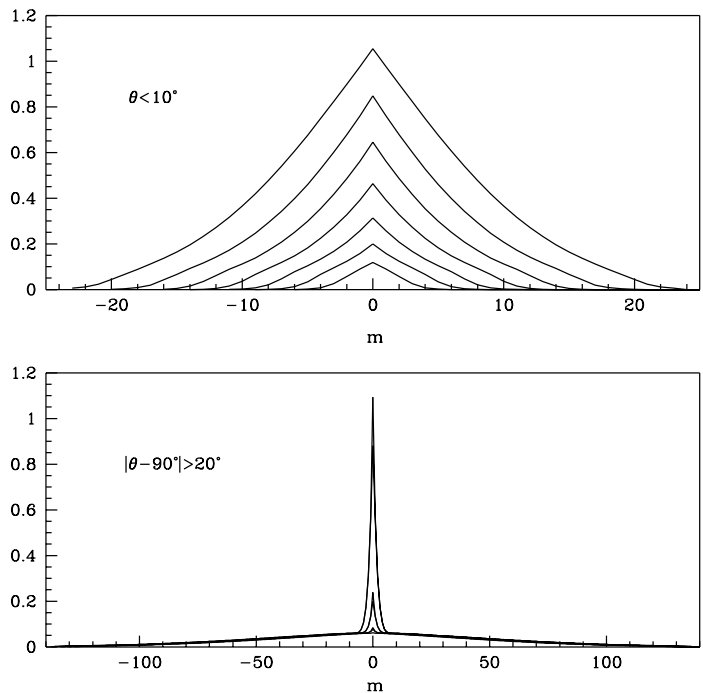,angle=0,width=9.5cm} 
\caption{The eigenvalues of $\mN^{-1/2}\mS\mN^{-1/2}$ at a given $m$ with
the tensor amplitude chosen to give a detection at $99\%$ confidence with
probability $0.5$ (see Sec.~\ref{sec:hyp}).
The noise is assumed isotropic and the model has reionization at $z=6.5$. 
For the small sky patch most of the signal is in the lowest few eigenmodes
of each $m$, but for larger patches a greater number of eigenmodes
are required to encompass all the signal (in the
bottom plot there are lots of contributions on top of each other along
the bottom line). For the large patch there are a small number of high
signal to noise modes due to the extra large scale reionization signal.\label{contribs}}
\end{center}
\end{figure}

In Fig.~\ref{contribs} we show the signal to noise in the magnetic
variables for two azimuthal patches.
As the patch size increases the signal in the modes with large $m$ also
increases, reflecting the fact that for small patches the diagonalization
of $\mW_+$ removes a greater relative fraction of the modes at each $m$ as
$m$ increases. For small patches of the
sky most of the signal at each $m$ is compressed into a small number of modes,
whereas for larger patches the signal is distributed more
uniformly. For cosmological models with reionization the signal for large patches
is distributed less uniformly, with a small number of modes giving
big contributions due to the greater large scale power.

\section{Measuring the magnetic signal}
\label{sec:hyp}

We are now in a position to use the magnetic observable $\vBw$ to
constrain the magnetic signal without having to worry about
contamination with the much larger electric signal.
The simplest thing to do would be to test the null hypothesis that the
magnetic signal is due entirely to noise (this hypothesis is unlikely to
be ruled out pre-Planck). If the signal were not consistent with noise it
could indicate various things: the presence of CMB magnetic
polarization, the presence of polarized foregrounds, that haven't been removed
successfully, systematic leakage into the magnetic mode in the analysis
(e.g.\ due to unaccounted for pointing errors, or pixelization effects),
or $Q$-$U$ leakage in the observation
(e.g.\ due to unaccounted for cross-polarization in the instrument
optics).

Magnetic polarization can originate from tensor modes, but also by
weak lensing of the scalar electric
polarization~\cite{Zaldarriaga98}. The lensing signal should be dominant on
small scales, and the magnetic variables could certainly be used to
observe this signal. Of more interest here is the larger scale contribution
from tensor modes. In order to identify this component we shall have
to model the lensing contribution, which becomes increasingly important as
one tries to observe smaller tensor contributions~\cite{Guzik00,Hu01}. In the
first three of the following subsections we assume that the magnetic
signal is generated purely from the tensor modes, then in
subsection~\ref{lensing} we show how our results can be adapted to
account for the lensing signal.

\subsection{Is it just noise?}

If the noise and signal are Gaussian the $\vBw$ will be
Gaussian and the simplest thing to do is a $\chi^2$ test by computing
$\chi^2 = \vBw^\dag \mN^{-1} \vBw$ (for isotropic noise this is just
$\chi^2 = \vBw^\dag \vBw/\sigma^2_N$). Whilst the CMB magnetic polarization
signal is from tensor modes is expected to be Gaussian, the lensing
signal and any spurious or unexpected signal
may not be. One may therefore also wish to do a more sophisticated set
of statistical tests at this point.

Assuming that the signal is as expected --- any $B$ signal
present is Gaussian and would have a power spectrum as predicted for a
near scale-invariant tensor initial power spectrum --- one can account for the expected form of
the power spectrum and thereby increase the chance of a detection. We assume
that the main parameters of the universe are well determined by the
time magnetic polarization comes to be observed, so the shape of the magnetic
polarization power spectrum is known to reasonable approximation
(the only significant freedom arising from the shape of the primordial tensor power
spectrum). We compute the expected signal correlation $\mS$ for some
particular tensor amplitude and say that the real signal is $r\mS$. 
Assuming Gaussian signal and noise the likelihood in this case is then
given by
\begin{equation}
L(\vBw|r) \propto \frac{ \exp[-\half \vBw^\dag (\mN+r\mS)^{-1}
\vBw]}{|\mN + r\mS|^{1/2}}.
\label{Likelihood}
\end{equation}
The likelihood distribution can be computed numerically from the
$\vBw$ observed, and gives the posterior probability distribution on
the value of $r$ after multiplying by the prior $f(r)$. 

The magnetic signal is expected to be weak, and the detailed
statistics for analysing such a signal are given in
Appendix~\ref{App:Stats}. There we show that
\begin{equation}
\nu'\equiv \frac{\vBw^\dag \mN^{-1} \mS
\mN^{-1} \vBw - \Tr(\mN^{-1}\mS)}{\sqrt{4 \vBw^\dag \mN^{-1}\mS\mN^{-1}\mS\mN^{-1}\vBw
- 2\Tr(\mN^{-1}\mS\mN^{-1}\mS) }}.
\end{equation}
gives a measure of the number of `sigmas' of the detection ---
the number of standard deviations of the maximum
likelihood $\hat{r}$ from pure noise ($r=0$) assuming low signal to noise. We use this as a test statistic in
Monte-Carlo simulations to compute detection probabilities at a given significance.   We have checked at
isolated points that using optimal statistics gains very little except for very small sky patches 
(where there are only a small number of magnetic modes,
each of which must have fairly high signal to noise in order to get a
detection).

Using the $\vBw$ variables is clearly not optimal as we have thrown
away some
well determined linear combinations of $E$ and $B$. However in
the idealized situation considered here they  should provide a robust
way for testing for magnetic polarization. The number of modes thrown
away is in any case quite small --- not more than two per $m$ mode
for azimuthal patches. We quantify this information loss further 
in Sec.~\ref{LOSSLESS}.

\subsection{Detection by Planck?}

\begin{figure}[t!]
\begin{center}
\psfig{figure=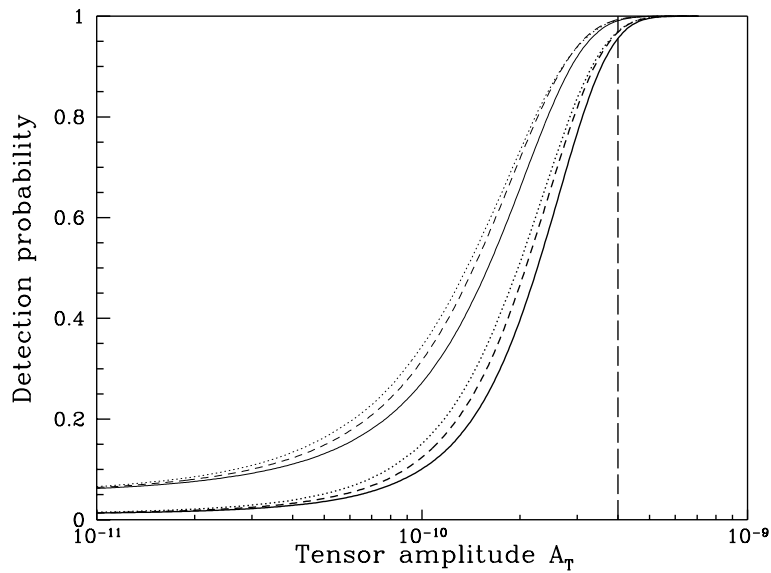,angle=-90,width=10.5cm} 
\caption{The probability of detecting magnetic polarization at 99 per cent
(thick lines)
and 95 per cent (thin lines) confidence as a function of the tensor initial
power spectrum amplitude $A_T$ for the model of Planck observations
described in the text. The dotted line is for a model with reionization at
$z=6.5$, using the unprojected variables and treating the electric
contamination as part of the noise (see Sec.~\ref{LOSSLESS}).  The other curves are using the
projected magnetic variables for models with no reionization (solid) and reionization
at $z=6.5$ (dashed). The vertical dashed line show the tensor amplitude that would
contribute about $1/10$ of the COBE signal.\label{planck_detect}}
\end{center}
\end{figure}

Of the current funded experiments, only Planck is likely to detect magnetic
polarization if the levels are as predicted by standard cosmological models.
As a toy model we consider the $143$ and $217$ GHz polarized channels of the
Planck High Frequency Instrument. We approximate the noise as isotropic
and ignore the variation of beam width (7.1 and 5.0 arcmin full width at half
maximum respectively) between these channels. Combining maps from these
two channels with inverse variance weighting, we find $\sigma_N \approx
6 \times 10^{-3} \mu \text{K}/\text{K}$, where $Q$ and $U$ are expressed
as dimensionless thermodynamic equivalent temperatures in units of the
CMB temperature. We apply an azimuthally-symmetric galactic cut of
$20$ degrees either side of the equator. The expected magnetic polarization
power spectrum peaks at $l\sim 100$, and there is therefore no
need to consider high resolutions so we can use $\lmax=250$ without
significant loss of power. In Fig.~\ref{planck_detect} we show the
probability of obtaining a detection with Planck as a function of the
true underlying scale-invariant tensor power spectrum amplitude $A_T$ (defined as in
Ref.~\cite{Martin00;long}) assuming a standard flat $\Lambda$CDM
model.

A tensor amplitude of $A_T\approx 4\times 10^{-10}$ would contribute about
1/10 of the large scale temperature $C_l$ detected by COBE, and
is likely to be detected by Planck if our model is at
all realistic. This corresponds to being able to detect the signal from inflationary models with
 energy scale at horizon crossing
$V^{1/4}\agt 2\times 10^{16}\,\text{GeV}$. Such models include the simple
$\phi^n$ potentials, with $n \ge 2$.

\subsection{Survey size}

For a given detector sensitivity the magnitude of the signal that
can be detected depends on the size of the sky patch that is observed.
The signal to noise in each observable increases in proportion to the
 observation time per unit area. The noise covariance
is proportional to $\sigma_N^2$ which varies in proportion to the
observed area for a given survey duration. For large areas the number of observables varies approximately in
proportion to the area, which would make the number of `sigmas' of a chi-squared
detection scale with the square root of the area. Combining these two effects,
the expected detection is therefore proportional to one over
the square root of the area, and is larger if a fixed survey time is
spent observing a smaller area. However for smaller areas
the signal to noise on each observable becomes larger, and the number
of variables decreases. With fewer variables the probability of
obtaining no detection increases significantly. This is just the
fact that if you observe a small patch of sky you have a larger chance
of being unlucky and having a patch which has a small magnetic
polarization signal everywhere. Also the existence of the boundary becomes
increasingly important for
small patches and a larger fraction of the information is lost in
order to obtain clean separation of the magnetic observables.

\begin{figure}[t!]
\begin{center}
\psfig{figure=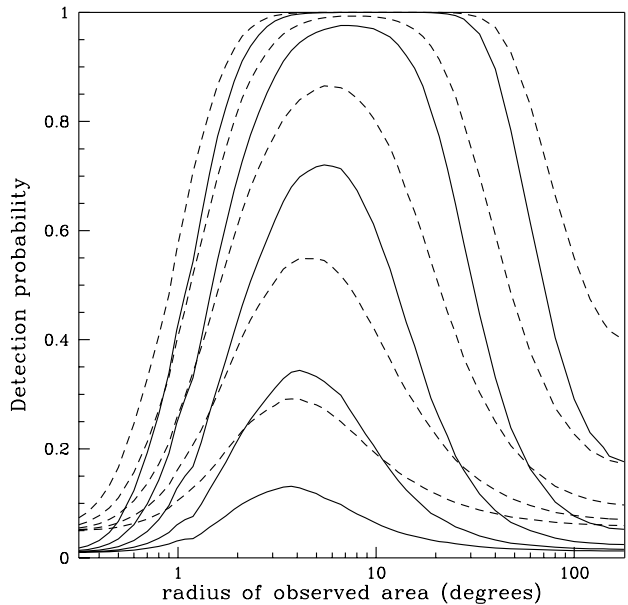,angle=0,width=8.5cm} 
\caption{The probability of being able to rule out the null hypothesis
at 95 per cent
(dashed lines) or 99 per cent (solid lines) confidence for scale invariant tensor amplitudes (bottom
to top) of $A_T= \{ 1,2,4,8,16\} \times 10^{-12}$, for a survey with detector sensitivity $s= 10\,\mu\text{K}
\sqrt{\text{sec}}$
that runs for one year and maps a circular patch of sky of a given
radius assuming uniform noise, no reionization, and no lensing. The probabilities
were computed by Monte-Carlo simulation.\label{patchsize_probs}}
\end{center}
\end{figure}

\begin{figure}[t!]
\begin{center}
\psfig{figure=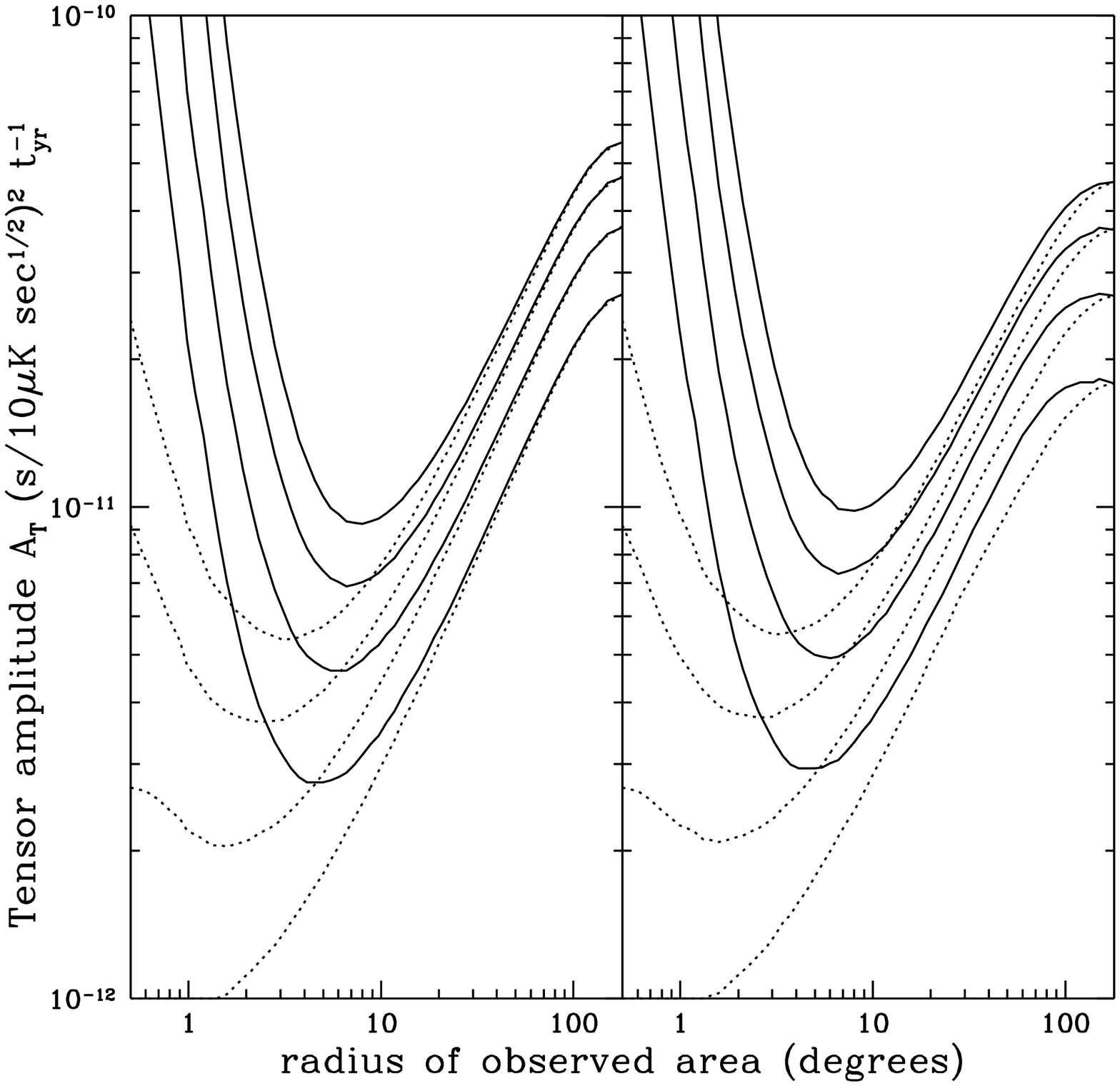,angle=0,width=12.5cm} 
\caption{The smallest gravitational wave amplitude $A_T$ (defined as
in Ref.~\cite{Martin00;long}) that could be detected at $99$ per cent
confidence with probability (bottom to top) of \{0.5,0.8,0.95,0.99\} by an
experiment with detector sensitivity $s= 10\,\mu\text{K}\sqrt{\text{sec}}$
that runs for one year and maps a circular patch of sky of a given
radius assuming uniform noise, no lensing, with reionization at
$z=6.5$ (right) and without reionization (left). 
The result scales with the square of the detector
sensitivity and inversely with the duration of the experiment. The
dotted lines show the equivalent result if one could perform perfect
lossless separation of the magnetic modes.\label{patchsize}}
\end{center}
\end{figure}

The question of `optimal'
survey size is somewhat delicate, as it depends on the probability
distribution for the detection significance that one thinks is optimal. In
Fig.~\ref{patchsize_probs} we plot the probability of detecting
various tensor amplitudes at 95 per cent and 99 per cent confidence for
different survey sizes. In Fig.~\ref{patchsize} we show the minimum gravitational wave (tensor)
amplitude that might be detected at 99 per cent confidence as a function of
the radius of the survey size. It is clear that radii in the range
$5^\circ$--$9^\circ$ are optimal, though one cannot be more precise
without defining more specifically the aims of the observation.
A radius of about $7^\circ$ would be a good
compromise between being able to place good upper limits if there was
no detection (which favours radii closer to $9^\circ$) and having a better
chance of detecting small amplitudes (which favours smaller radii).

The solid curves in Fig.~\ref{patchsize} fully take account of the need
to separate the magnetic signal from the (much larger) electric signal.
By way of comparison, the dashed curves show the minimum detectable amplitude
obtainable if one could do perfect lossless separation, which is
clearly impossible on an incomplete sky (see Sec.~\ref{LOSSLESS}). With
lossless separation, the best upper bounds are obtained for smaller patches
since the size of the boundary is no longer important.
The dashed curves in Fig.~\ref{patchsize} can be compared with those given in 
Ref.~\cite{Jaffe00a} where perfect separation was assumed, the effects of
finite sky coverage were treated only approximately, and a less rigorous
approach to hypothesis testing was employed. Ref.~\cite{Bunn01}
gives an improved analysis along the lines of Ref.~\cite{Jaffe00a}, and
also performs calculations properly taking account of the mixing of
electric and magnetic polarization through a (brute-force) Fisher
analysis in pixel space.

\subsection{Lensing}
\label{lensing}

Unlike most of the foreground signals that might contaminate the observation,
the magnetic signal from the lensing of scalar electric polarization
has the same frequency spectrum as the primordial magnetic signal
and so cannot be removed easily by use of multi-frequency
observations. In order to isolate the tensor contribution to the
magnetic signal we can incorporate knowledge of the expected lensing
power spectrum~\cite{Zaldarriaga98} into the null-hypothesis
covariance matrix $\mN$ (we neglect the non-Gaussianity of the lensed polarization). For the multipoles $l \alt 250$ of interest for the tensor
signal the lensing signal is approximately white, with
$C^{BB}_{\text{lens}} \approx 4.4\times 10^{-6} \mu \text{K}^2$ if the COBE
signal is entirely generated by scalar modes. For large patches of sky,
where the matrix $\mDt_+$ is nearly proportional to the identity
matrix, the lensing signal contributes like an additional isotropic noise with
$\sigma_{N,\text{lens}}^2\approx C^{BB}_{\text{lens}}$. We have
checked this approximation by computing the following results exactly in
particular cases, with agreement to within a few percent for patch
sizes with $10^\circ \alt \theta \alt 80^\circ$.

\begin{figure}[t!]
\begin{center}
\psfig{figure=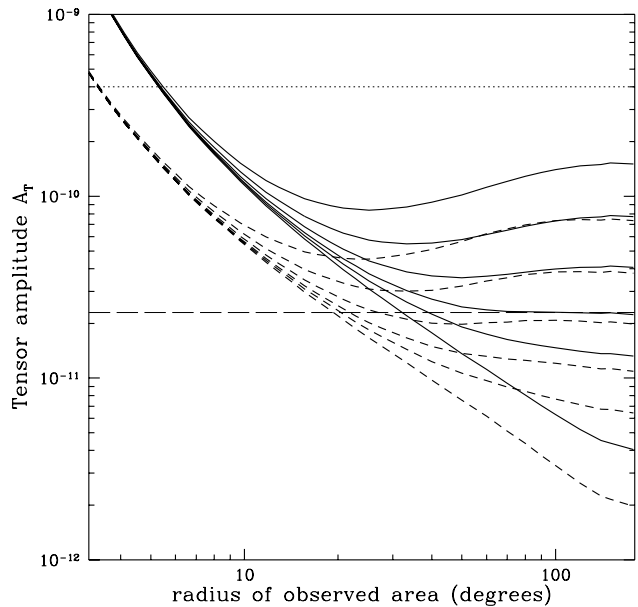,angle=0,width=8.5cm} 
\caption{The smallest gravitational wave amplitude $A_T$ that could be detected at $99$ per cent
confidence with probability $0.95$ (solid lines) and $0.5$ (dashed lines) by an
experiment with detector sensitivity (bottom to top) $s^2= \{0, 25, 50,
100, 200, 400\}\,\mu\text{K}^2\text{sec}$
that runs for year and maps a circular patch of sky of a given
radius assuming uniform noise, a white lensing power spectrum, and a
$\Lambda$CDM cosmology that reionizes at $z=6.5$. 
The horizontal dotted line shows the amplitude which contributes about
$1/10$ of the COBE signal. The long dashed line shows very
roughly the dust foreground at $143\,\text{GHz}$. \label{patchsize_lens}}
\end{center}
\end{figure}

The effect of the lensing is therefore simply
to increase the effective noise by a constant amount. For the Planck satellite
the effect is small, reducing the $A_T$ that could be observed by
about $1.5$ per cent. However for the smaller surveys with better sensitivity,
considered in Figs.~\ref{patchsize_probs} and~\ref{patchsize}, the effect
is much more important. For a one year survey of radius $\theta$
with sensitivity $s =10\,\mu\text{K}\sqrt{\text{sec}}$ the noise
variance is given by $f(\theta)\sigma^2_0$ where
$f(\theta)=(1-\cos\theta)/2$ is the fraction of the sky which is
observed and $\sigma_0^2 \approx 4\times 10^{-5} \mu\text{K}^2$.
This noise gives the tensor amplitudes $A_T(\theta)$ plotted in
Fig.~\ref{patchsize}. Incorporating the lensing effect the actual tensor
amplitude one could detect in an experiment with sensitivity $s$ and duration
$T$ is
\begin{equation}
A_T(\theta)' = A_T(\theta) \frac{4\pi f(\theta)s^2/T +\sigma_{N,\text{lens}}^2}{f(\theta)\sigma^2_0},
\end{equation}
where $A_T(\theta)$ is the amplitude for a one year mission with
$s =10\,\mu\text{K}\sqrt{\text{sec}}$ and lensing ignored (i.e.\ the
amplitude plotted in Fig.~\ref{patchsize}).
This allows our previous results to be modified for inclusion of the
lensing signal. We have plotted the modified results
for various survey sensitivities in Fig.~\ref{patchsize_lens}. The optimal
survey size now depends on the sensitivity---as sensitivity improves
the lensing signal becomes more important and one needs to survey
larger scales in order to accurately measure the difference in
variance expected with the tensor signal.
For large patch sizes the tensor amplitude $A_T$ that can be detected
in the absence of lensing is proportional to $\sqrt{f(\theta)}$. Allowing
for lensing, there is an optimal survey size at $4\pi s^2 f(\theta)/T =
\sigma_{N,\text{lens}}^2$ [if there is a solution with $f(\theta)<1$],
in other words when the variance of the instrument noise is equal to the
lensing signal. There is a lower limit of
$A_T\approx 4\times 10^{-12}$ that can be measured even with perfect
sensitivity, when the tensor contribution cannot be distinguished from
random sampling variations in the lensing
signal distribution. This corresponds to an inflation model with
energy scale $V^{1/4}\approx 7\times 10^{15}\, \text{GeV}$, in broad
agreement with Ref.~\cite{Hu01} for a three sigma detection. This situation
could only be improved if one could find ways to obtain information about
the particular realization of the lensed signal.

We have assumed that component separation and source subtraction can
be performed exactly so that the observed signal is only lensed CMB.
Polarized thermal dust emission is expected to generate a
significant magnetic signal~\cite{Prunet99} at roughly the level shown
by the line in Fig.~\ref{patchsize_lens} at $143 \, \text{GHz}$.
Separation of this signal from the CMB
signal should be possible with multi-frequency observations, and it
should then not have a significant effect on our results.

\section{Lossless methods}
\label{LOSSLESS}

We now compare the above analysis with truly lossless methods.
Lossless, likelihood analysis for CMB polarization in pixel space has been
considered recently in Ref.~\cite{Bunn01}. In this section we consider
lossless and nearly lossless methods in harmonic space.

A simple way to incorporate most of the magnetic signal for constraining the
tensor modes is to use the unprojected variables
$\mDt_+^{-1/2}\mUt^\dagger_+\vBt$, where $\mDt_+$ and $\mUt_+$ were
defined in Sec.~\ref{sec:sep}. The
null-hypothesis covariance matrix $\mN$ can be computed including the
expected signal from the electric polarization, and the analysis can
be performed as before. This is marginally superior to using the
projected variables if the tensor amplitude is quite high, as shown in Fig.~\ref{planck_detect}
for the Planck mission. For smaller tensor amplitudes the entangled linear combinations of $E$ and $B$ modes are dominated
by the electric component and performing the projection looses very
little. Using the projection gives one clean separation, and there is
no need to know the electric polarization power spectrum. By identifying
variables that depend only on the electric and magnetic signals at the
level of the map we also do not need to assume Gaussianity,
so we could for example perform Gaussianity tests on the two
physically distinct polarization types independently.

We now consider the full joint analysis of the electric and magnetic
polarization, with the pseudo-multipoles
$\vEt$ and $\vBt$ forming our fundamental data vector
\begin{equation}
\begm \vEt \\ \vBt\enm = \begm \mW_+ & i\mW_- \\ -i\mW_- & \mW_+\enm
\begm \vE \\ \vB \enm.
\end{equation}
Since we are no longer worrying about $E$-$B$ separation so we can
equally well use
the block-diagonal frame where
\begin{equation}
\vAt \equiv \begm \vEt + i\vBt \\
\vEt -i\vBt\enm = \begm {}_2\mW & \mzero \\ \mzero &
{}_{-2}\mW \enm
\begm \vE + i\vB \\ \vE-i\vB \enm \equiv
\begm {}_2\mW & \mzero \\ \mzero &
{}_{-2}\mW \enm \vA .
\end{equation}
Performing a singular value decomposition of the block-diagonal matrix
$\text{diag}({}_2\mW,{}_{-2}\mW) = \mU \mD \mU^\dagger$ (where the
matrices $\mU$ and $\mD$ should not be confused with those defined
in Sec.~\ref{sec:sep}),
we can identify the well determined linear combinations as before
\begin{equation}
\vA' \equiv  \mDt^{-1/2} \mUt^\dagger \vAt \approx
\mDt^{1/2} \mUt^\dagger \vA.
\end{equation}
This diagonalization is equivalent to 
defining harmonic coefficients $\vA'$ with respect to a complete set of
spin two harmonics which are orthonormal over the patch of sky, in the
same way as for the spin zero cut-sky temperature
analysis~\cite{Gorski94,Mortlock00}. As before, this construction
ensures that for isotropic noise the noise correlation is diagonal
\begin{equation}
\la \Delta \vAt \Delta \vAt{}^\dag \ra =2 \sigma^2_N
\text{diag}({}_2\mW,{}_{-2}\mW) \quad\implies\quad 
\mN \equiv \la \Delta \vA' \Delta \vA'{}^\dag \ra = 2\sigma^2_N \mI.
\end{equation}
The signal correlation is given by
\begin{equation}
\mS = \la \vA'\vA'{}^\dag \ra = \mDt^{1/2} \mUt^\dagger \begm \mC^{EE} + \mC^{BB} &
\mC^{EE} - \mC^{BB} \\ \mC^{EE} - \mC^{BB} & \mC^{EE} + \mC^{BB} \enm
\mUt \mDt^{1/2},
\end{equation}
where the $\mC^{EE}$ and $\mC^{BB}$ are the diagonal electric and
magnetic power spectrum matrices respectively, and we have assumed
that $\mC^{EB}=0$. If the noise and signal are Gaussian
we can proceed to do a likelihood analysis for the power
spectra using
\begin{equation}
L(\vA'|\mC^{EE},\mC^{BB}) \propto \frac{\exp [-\half \vA'{}^\dag
(\mN+\mS)^{-1} \vA']}{|\mN+\mS|^{1/2}}.
\end{equation}
The coupling matrices ${}_{\pm 2}\mW$ can be computed quickly for an
azimuthally symmetric patch of sky, as described in
Appendix~\ref{App:Ints}, and modes with different $m$ decouple. The
problem is therefore tractable. However it is not nearly so simple to
find a maximum likelihood estimate of the magnetic amplitude, and in
general there will be complicated correlations between the recovered
power spectra. By using the lossy projection in the previous sections
we have essentially shown that this likelihood function is `nearly'
separable. Making it separable costs something in terms of lost
information, but it significantly simplifies the problem. 
 Using the
projected variables also reduces the size of the matrices, so
performing the matrix inversions is significantly faster. 

If the $B$ signal is determined to be negligible one would want to
apply an efficient nearly-lossless method  to estimate the electric power
spectrum $C^{EE}_l$ (or to do parameter estimation), so we now consider the case when one polarization type is absent. If
the $B$ signal can be neglected we have
\begin{equation}
\begm \vEt \\ i \vBt\enm \approx \begm \mW_+  \\ \mW_- \enm \vE = \mU \mD
\mV^\dagger \vE,
\end{equation}
where we have done a singular value decomposition as
before so that we can find the well determined linear combinations of
the $E_{lm}$:
\begin{equation}
\vx_E \equiv \mDt^{-1/2} \mUt \begm \vEt \\ i \vBt\enm \approx
\mDt^{1/2} \mVt^\dagger \vE.
\end{equation}
The matrices one has to invert to do a likelihood analysis are now one
half the size of those in the optimal method when both
polarization types are present, and so the problem is numerically much
faster. However isotropic noise no longer gives the simple diagonal
noise covariance, though this can always be rectified by using $\mN^{-1/2}\vx_E$. In practice a nearly optimal method would probably
be more appropriate using only the unprojected variables
$\mDt_+^{-1/2}\mUt^\dagger_+\vEt$, where $\mDt_+$ and $\mUt_+$ were defined
in Sec.~\ref{sec:sep}. These variables have diagonal noise
properties like the $\vEw$ for isotropic noise, and the computational
saving may be significant when analysing high resolution polarization
maps. We have checked numerically that including $\vBt$ in the
analysis  gains very little even for low
$\lmax$ and small patches. For large area surveys at high resolution
the information loss will probably be negligible.

There are exactly equivalent relations for the well determined
magnetic variables in the case when $E$ vanishes. 
This case is of little practical interest, since the $E$ signal would
have to be removed to within the magnitude of the $B$ signal, and this
is impossible on an incomplete sky since the two are not unambiguously
distinguishable without accurate boundary data. However, supposing that $E$
could be removed is useful theoretically as we can then compute the best
obtainable magnetic signal to compare
with what we obtain using our projected variables. The information
lost due to the projection depends on the cosmology. Models with reionization
have more power on large scales and a greater fraction of the power is lost due
to removal of the boundary terms. For our toy model of the Planck satellite
we find that the amplitude that could be detected at given
significance and probability is reduced by  about
$30$ per cent by the projection for a cosmology with
reionization at $z=6.5$, but only by $2$ per cent for a zero reionization
model. In the reionization model one is loosing a lot of the additional
information in the low multipoles that in the absence of the
projection would have high signal to noise. The net result is that the
reionization model has an only slightly higher chance of giving a
tensor detection despite having more large scale power. By using the unprojected variables and incorporating
the expected electric polarization contamination as an extra noise term one
can approximately halve this loss. 
The lossless result
is compared to the realistic projected result for general
circular sky patches in Fig.~\ref{patchsize} for an observation with
much higher sensitivity.
The cost we incur by using the non-optimal method in terms of slightly larger
error bars on the $B$ signal, or a less powerful test of detection at a given
significance, is small for large survey
areas though it does increase for small sky patches. For these
sensitive observations the electric signal is much larger than the
magnetic signal and essentially nothing is lost by performing the
projection rather than including electric contamination as a large extra noise.
 
To make a detection of the magnetic signal on such a
small sky patch with the planned long duration balloon observations the tensor/scalar ratio would need to be significantly
larger than one, which is too large
to be allowed by the current temperature anisotropy
observations~\cite{Wang01}. Of course, seeing if there is only a small
magnetic signal is an important consistency check for current models
with low tensor/scalar ratio to pass. 

One simple way to reduce the information loss in our method would
be to use data objects that include not only the surface integrals
$I_{\pm 2,W}$, but also those parts of the  boundary terms in
Eqs.~(\ref{eq:i2wp}) and (\ref{eq:im2wp}) that do not depend on
$\edth P$ on the boundary. Such objects would separate electric and magnetic
polarization exactly if the scalar window functions were constructed to
vanish on the boundary. The problem of producing a non-redundant set of
such windows could be tackled with a simple variant of the harmonic-based
method presented in Sec.~\ref{sec:sep}. The
additional boundary contribution would cancel that part of
$\mW_-$ that couples to the normal derivative of the window function
on the boundary, leaving a single non-zero singular value (for $|m|>0$) to
project out. The net effect would be that for azimuthal
patches we would gain one extra variable per $m$ for $|m| > 1$, though
the noise properties of these variables would not be as simple as if the
line integrals were not included, and the problem of performing line
integrals with pixelized data is non-trivial. For
reionization models (which have significant large scale power)
the reduction in information loss may be worth the effort
required to overcome these obstacles, though a full analysis with the
non-separated variables would probably work better.

\section{Conclusion}

We have considered the problem of producing statistically independent
measures of the electric and magnetic polarization from observations covering
only a portion of the sky. Although the separation of the polarization field
into electric and magnetic modes is not unique in the presence of
boundaries, we have shown how to construct window functions that are
guaranteed to probe separately the electric and magnetic polarization exactly
over azimuthally-symmetric patches of the sky. We presented a harmonic-based
method for efficient construction of the windows that automatically removes
redundancy due
to the finite sky coverage. In addition, our window functions return separated
electric and magnetic variables that have very simple diagonal noise
correlations for idealized noise on the polarization map.
For azimuthal patches separating the electric and magnetic
polarization comes at the cost of losing two pieces of information
per $m$ mode, or roughly twice the number of pixels of area $(\pi/\lmax)^2$
on the boundary of the patch. For large patches this information loss is
small unless there is large scale power due to reionization, but for smaller patches it can be more severe due to the limited
support of the high $m$ spin-weight 2 harmonics in the patch.
Although we have proved that our method gives exact separation for
azimuthal patches, the harmonic-based construction should produce
window functions that give approximate separation for arbitrarily shaped
patches with similar information loss to the azimuthal case.

We showed how the variables constructed from our window functions
could be used to constrain the amplitude of the magnetic signal without
contamination from the much larger electric signal. For the first time,
we made predictions for the tensor amplitude that Planck should be able
to detect taking proper account of excluding the galactic region. If other
non-negligible foregrounds can be removed using the other frequency
channels, Planck should be able to detect the magnetic signal predicted
by some simple inflationary models. For less sensitive observations, our
window functions should nevertheless be useful to set upper
limits on the magnetic signal, and may also aid the identification
of systematic effects in the instrument or analysis pipeline.

If the magnetic signal is shown to be consistent with noise
then we showed how one can use all the well determined polarization
pseudo-multipoles to analyse the electric polarization power spectrum without
loss of information.  The analysis using these variables is no more
complicated than the analysis of temperature anisotropies using cut-sky
orthogonalized scalar harmonic functions~\cite{Gorski94,Mortlock00}.

We have only considered isotropic noise here, however, as long as the
noise is azimuthally symmetric the separation of $m$ modes will still
work, and the problem remains computationally tractable though rather
less simple. In practice, there will be several other complications in
real-life CMB polarimetry observations that will impact on the map-making
and subsequent analysis stages. Further careful investigation of the
propagation of instrument effects such as beam asymmetries,
straylight, cross-polar contamination, and pointing instabilities through the
map-making stage will be required before the programme for analysing
magnetic polarization outlined in this paper will be realizable.

\begin{acknowledgments}
We acknowledge use of the LAPACK
package\footnote{\link{http://www.netlib.org/lapack/}}
for performing the matrix decompositions. We acknowledge the support
of PPARC via a  PPTC
Special Program Grant for this work.
AC acknowledges a PPARC Postdoctoral Fellowship. We also thank
PPARC and HEFCE for support of the COSMOS facility.
\end{acknowledgments}

\appendix

\section{Spin raising and lowering operators}
\label{App:Eth}

In general a spin-weight $s$ quantity ${}_s\eta$ is defined over a
two-dimensional Riemannian manifold with respect to an orthonormal diad field
$\{\ve_1,\ve_2\}$. The local freedom in the choice of diad amounts to the
transformations
\begin{equation}
\ve_{\pm} \equiv \ve_1 \pm i\ve_2 \rightarrow e^{i\gamma} \ve_{\pm}
\label{eq:appa0a}
\end{equation}
of the (complex) 
null vectors $\ve_\pm$. A quantity is defined to be of spin-weight
$s$ if under the transformation (\ref{eq:appa0a}) ${}_s\eta \rightarrow
{}_s\eta e^{is \gamma}$.
To every spin-weight $s$ object ${}_s\eta$ we can associate a (complex)
symmetric trace-free, rank-$|s|$ tensor $\eta_{A_s} \equiv
\eta_{a_1 \dots a_s}$: for $s\geq 0$,
\begin{equation}
\eta^{A_s} \equiv 2^{-s} {}_s \eta e_-^{A_s},
\end{equation}
where the irreducible tensor product $e_-^{A_s} \equiv e_-^{a_1} \dots
e_-^{a_s}$. The inverse relation is
\begin{equation}
{}_s \eta = \eta_{A_s} e_+^{A_s}.
\end{equation}
For $s<0$ we define $\eta^{A_{|s|}} \equiv 2^{-|s|} {}_s\eta e_+^{A_{|s|}}$.
The spin raising and lowering operators $\edth$ and $\edth$ are defined by
the null diad components of the covariant derivatives of $\eta_{A_{|s|}}$:
\begin{eqnarray}
\edth {}_{\pm |s|}\eta &=& - (e_+^c \nabla_c \eta_{A_{|s|}}) e_{\pm}^{A_{|s|}},
\label{eq:appa1c}\\
\beth {}_{\pm |s|}\eta &=& - (e_-^c \nabla_c \eta_{A_{|s|}}) e_{\pm}^{A_{|s|}}.
\label{eq:appa1d}
\end{eqnarray}
(The minus signs are conventional).

In CMB polarimetry we are concerned with fields defined over the sphere,
in which case the transformation in Eq.~(\ref{eq:appa0a}) corresponds to
a \emph{left}-handed rotation of the diad about the outward normal
$\hat{\mathbf{r}}$. Choosing the orthonormal diad to be aligned with the
coordinate
basis vectors $\hat{\btheta}$ and $\hat{\bphi}$ of a spherical polar coordinate
system, we have $e_\pm^a \nabla_a e_\pm^b = \cot\theta e_\pm^b$ and
$e_\mp^a \nabla_a e_\pm^b = - \cot\theta e_\pm^b$. It follows that for
this choice of diad the spin raising and lowering operators reduce to
\begin{eqnarray}
\edth {}_s\eta &=& - \sin^s\theta(\partial_\theta + i \csc\theta
\partial_\phi)(\sin^{-s}\theta {}_s\eta), \\
\beth {}_s\eta &=& - \sin^{-s}\theta(\partial_\theta -i \csc\theta
\partial_\phi)(\sin^{s}\theta {}_s\eta).
\end{eqnarray}

An elegant interpretation of the spin raising and lowering operators on the
sphere can be obtained by considering spin-weight $s$ objects
${}_s\eta(\theta,\phi,\psi)$ defined on a diad at (position-dependent)
angle $\psi$ to the coordinate directions, so that
\begin{equation}
{}_s\eta(\theta,\phi,\psi) = {}_s\eta(\theta,\phi)e^{is\psi},
\end{equation}
where ${}_s\eta(\theta,\phi)$ is defined on $\hat{\btheta}$ and $\hat{\bphi}$.
In this case, the spin raising and lowering operators can be related to
the angular momentum operators for a rigid body~\cite{Goldberg67}.
Working in a representation where the orientation of the body is specified in
terms of Euler angles $(\phi,\theta,-\psi)$\footnote{Our convention
for Euler angles $(\alpha,\beta,\gamma)$ follows Ref.~\cite{Brink93}, i.e.,
successive right-handed rotations by $\gamma$, $\beta$, and $\alpha$ about
the $z$, $y$, and $z$-axes respectively. The use of $\psi$, which is minus
the third Euler angle, as a configuration variable for the rigid body is
necessary to relate the angular momentum operators directly to the
spin raising and lowering operators with the (consistent)
conventions we have adopted here.}, the angular momentum operators
on the \emph{body-fixed axes} take the form
\begin{eqnarray}
K_z     &=& i \partial_\psi, \\
K_{\pm} &=& e^{\pm i \psi}(\pm \partial_\theta + i \csc\theta \partial_\phi
+ i \cot\theta \partial_\psi),
\end{eqnarray}
where $K_{\pm} \equiv K_x \pm i K_y$. These operators satisfy the commutation
relations
\begin{equation}
[K_z,K_\pm] = \mp K_\pm, \qquad [K_+,K_-]=-2 K_z,
\label{eq:appa0}
\end{equation}
so that $K_\pm$ are lowering/raising operators with respect to the eigenvalues
of $K_z$. Note that the signs in these commutation relations are different
from those for angular momentum operators on a fixed frame since on the
body-fixed axes we have $[K_x,K_y] = - i K_z$~\cite{LandauLifshitzQM}.
The action of the spin raising and lowering operators can then be
formulated in terms of the angular momentum operators as
\begin{eqnarray}
K_+ {}_s\eta(\theta,\phi,\psi)&=&- e^{i(s+1)\psi} \edth{}_s\eta(\theta,\phi),
\label{eq:appa1} \\
K_- {}_s\eta(\theta,\phi,\psi)&=&+ e^{i(s-1)\psi} \beth{}_s\eta(\theta,\phi).
\label{eq:appa2}
\end{eqnarray}

Several useful results for the spherical raising and lowering operators
follow from the commutation relations (\ref{eq:appa0}). [Similar relations
on a general manifold can be found from Eqs.~(\ref{eq:appa1c}) and
(\ref{eq:appa1d}).] For a spin-weight $s$ quantity defined on the
coordinate basis,
\begin{eqnarray}
(\beth\edth-\edth\beth){}_s\eta(\theta,\phi) &=& 2s {}_s\eta(\theta,\phi),\\
{[} \edth \beth - s(s-1)]
{}_s\eta(\theta,\phi) &=& (\nabla^2 - s^2 \csc^2\theta
+ 2i s \cot\theta\, \csc\theta\, \partial_\phi) {}_s\eta(\theta,\phi),
\end{eqnarray}
where we have used $K_+ K_- = K^2 - K_z - K_z^2$ to derive the last identity.
Applying these relations repeatedly to a spin-weight 0 quantity we find
\begin{equation}
\edth\edth\beth\beth {}_0\eta(\theta,\phi) =
\beth\beth\edth\edth {}_0\eta(\theta,\phi) =
(\grad^2+2)\grad^2 {}_0\eta(\theta,\phi),
\label{ethethbethbeth}
\end{equation}
and the useful relation
\begin{equation}
(\beth\edth^{\!s} - \edth^{\!s}\beth) {}_0\eta(\theta,\phi) =
s(s-1)\edth^{\!s-1} {}_0\eta(\theta,\phi).
\label{commident}
\end{equation}

\subsection*{Integral theorems}

The integral of the spin-weight 0 quantity $\edth {}_{-1}\eta$ over some
portion $S$ of the two-dimensional manifold is determined by the
integral around the boundary $\partial S$:
\begin{equation}
\int_{S}  \text{d}S \,\edth {}_{-1}\eta = \oint_{\partial S}
\deth  {}_{-1} \eta,
\label{inttheorem}
\end{equation}
where $\deth $ is the spin-one element of length around the boundary:
\begin{equation}
\deth  \equiv i \text{d}l_a e_+^a.
\end{equation}
On the sphere in the spherical polar coordinate frame  $\deth  = i \text{d}\theta -
\sin\theta\, \text{d}\phi$. Equation~(\ref{inttheorem}) is the complex
representation of Stokes' Theorem and the Divergence Theorem. An equivalent
result holds for spin-weight one quantities by forming the complex conjugate
of the above, with $\dbeth = -i \text{d} l_a e_-^a$.

The spin raising and lowering operators obey Leibnitz' rule, so there
is a `Green's  identity'
\begin{equation}
P\edth\beth Q - Q\beth\edth P = \edth(P\beth Q) - \beth(Q\edth P),
\end{equation}
where $P$ and $Q$ have definite spin weight. For $PQ$ with spin-weight zero
integrating over a surface gives the integral theorem
\begin{equation}
\int_{S} \text{d}S (P\edth\beth Q - Q\beth\edth P) = \oint_{\partial S}
\deth P\beth Q - \oint_{\partial S} \dbeth Q\edth P.
\label{inttheorem1}
\end{equation}
A similar result is obtained using
\begin{equation}
P\edth\edth Q - Q\edth\edth P = \edth( P\edth Q - Q\edth P)
\end{equation}
for $PQ$ of spin weight $-2$ which gives
\begin{equation}
\int_{S} \text{d}S (P\edth\edth Q - Q\edth\edth P) = \oint_{\partial S}
\deth ( P\edth Q - Q \edth P),
\label{inttheorem2}
\end{equation}
with a similar result  for a spin-weight $2$ quantity.

\section{Spin weighted spherical harmonics}
\label{App:harmonics}

The spin-weight $s$ spherical harmonics ${}_s Y_{lm}$ are defined in terms of
the usual spherical harmonics $Y_{lm}$ by
\begin{equation}
{}_sY_{lm} \equiv \sqrt{\frac{(l-|s|)!}{(l+|s|)!}}\,\, \edth^{\!s}\, Y_{lm},
\label{eq:appbharm}
\end{equation}
where $\edth^{-|s|}\equiv (-1)^s \beth^{|s|}$. They are non-zero for
$|s|\le l, |m|\le l$. By making use of Eqs.~(\ref{eq:appa1}) and
(\ref{eq:appa2}), and the properties of the $K_{\pm}$ operators when acting
on Wigner $D$-matrices\footnote{Our conventions for the $D$-matrices follow
Refs.~\cite{Brink93,AngularMom}. We adopt the Condon-Shortley phase for
the spherical harmonics, which differs from that used by Goldberg
et al.~\cite{Goldberg67} by a factor of $(-1)^m$.} \cite{Brink93,AngularMom},
it is straightforward to show that~\cite{Challinor00}
\begin{equation}
D^l_{-m s}(\phi,\theta,-\psi) = (-1)^m \sqrt{\frac{4\pi}{2l+1}}
{}_s Y_{lm}(\theta,\phi) e^{is\psi}.
\label{eq:appb1}
\end{equation}
With the conventions adopted here, ${}_sY_{lm}^\ast = (-1)^{s+m}{}_{-s}Y_{l(-m)}$.
Orthonormality of the spin weight harmonics over the full sphere,
\begin{equation}
\int_{4\pi} \text{d}S \, {}_s Y_{lm} {}_s Y_{l' m'}^\ast = \delta_{ll'}
\delta_{mm'},
\end{equation}
follows from the orthogonality of the $D$-matrices over the $\text{SO}(3)$
group manifold.

Applying $K_+ K_-$ to Eq.~(\ref{eq:appb1}), and using Eqs.~(\ref{eq:appa1}) and
(\ref{eq:appa2}), one can show that the spin weight harmonics satisfy the
differential equation
\begin{equation}
\edth\beth\, {}_sY_{lm} = \left[s(s-1)-l(l+1)\right]\, {}_sY_{lm}.
\label{eq:appb4}
\end{equation}
[An alternative proof of this result follows from Eq.~(\ref{commident}).]
The spin weighted harmonics are separable and can be written as
\begin{equation}
{}_sY_{lm}(\theta,\phi) = {}_s\lambda_{lm}(\cos\theta) e^{im\phi}.
\end{equation}
The ${}_s\lambda_{lm}$ satisfy the self-adjoint equation
\begin{equation}
[(1-x^2) {}_s\lambda_{lm} ']' - \frac{m^2+s^2+2 m s x}{1-x^2}
{}_s\lambda_{lm} = -l(l+1){}_s\lambda_{lm},
\label{diffequation}
\end{equation}
where a prime denotes differentiation with respect to $x = \cos\theta$.

The ${}_s\lambda_{lm}$ can be evaluated recursively for $m \geq |s|$ starting
with
\begin{equation}
{}_s\lambda_{mm}(x) = (-2)^{-m}\sqrt\frac{(2m+1)!}{4\pi(m+s)!(m-s)!}
(1-x)^{(m+s)/2} (1+x)^{(m-s)/2},
\label{eq:appc5}
\end{equation}
and the recursion relation (derived from standard
results for the Wigner $D$-matrices; see e.g.\ Ref.~\cite{AngularMom})
\begin{equation}
{}_s\lambda_{lm} = \left( x + \frac{s m}{l(l-1)}\right) C_{slm}{}_s\lambda_{(l-1)m} - \frac{ C_{slm}}{
C_{s(l-1)m}} {}_s\lambda_{(l-2)m},
\label{Lambdarecur}
\end{equation}
where
\begin{equation}
C_{slm} \equiv \sqrt\frac{l^2(4l^2-1)}{(l^2-m^2)(l^2-s^2)}.
\end{equation}
The harmonics for $m \leq - |s|$ can be obtained from ${}_s\lambda_{l(-m)}=
(-1)^{s+m} {}_{-s}\lambda_{lm}$. A straightforward way to evaluate
${}_s\lambda_{lm}$ for $|m| < |s|$ is to compute ${}_{\pm n} \lambda_{l |s|}$
for $0 \leq n < |s|$ and then use the symmetry
${}_s\lambda_{lm} = (-1)^{m+s} {}_m\lambda_{ls}$.
Further useful results for the spin weight harmonics can be found
in Refs.~\cite{Zaldarriaga97,Ng99}, and expressions for the spin-weight $\pm2$
harmonics in terms of the associated Legendre functions are given
in Ref.~\cite{Kamionkowski97}.

\section{Overlap Integrals}
\label{App:Ints}

Setting $P={}_sY_{l'm'}^\ast$ and $Q={}_sY_{lm}$ in Eq.~\eqref{inttheorem1},
and using the differential equation~(\ref{eq:appb4}) to simplify the
integrand on the left-hand side, we have
\begin{eqnarray}
[l'(l'+1) - l(l+1)]\int_{S} \text{d}S\, {}_sY_{l'm'}^\ast\, {}_sY_{lm}
= \oint_{\partial S} \deth {}_sY_{l'm'}^\ast \,\beth {}_sY_{lm} -
\oint_{\partial S} \dbeth {}_sY_{lm}\edth {}_sY_{l'm'}^\ast.
\label{eq:appc1}
\end{eqnarray}
This expresses the overlap integral for $l\ne l'$ in terms of a line
integral around the boundary of $S$.

For an azimuthally symmetric surface the integrals are particularly
straightforward as the spin-weight harmonics with different $m$ are
orthogonal over the patch. The overlap integral for the same $m$ but
different $l$ can be determined from Eq.~\eqref{eq:appc1} to be
\begin{equation}
[l(l+1)-l'(l'+1)]\int_a^b \text{d}x\,{}_s\lambda_{l'm}\, {}_s\lambda_{lm} = 
(1-x^2)[ {}_s\lambda_{l'm}' {}_s\lambda_{lm}-{}_s\lambda_{lm}'
{}_s\lambda_{l'm}]_a^b.
\end{equation}
[This result also follows directly from Eq.~\eqref{diffequation}.]
The derivatives can be removed while maintaining homogeneity in the
spin weight by using
\begin{equation}
l(1-x^2) {}_s\lambda_{lm}' = -(sm + l^2 x) {}_s\lambda_{lm} +
\frac{(2l+1)l}{C_{slm}} {}_s\lambda_{(l-1)m}
\end{equation}
to write the $l'\ne l$ integral as
\begin{eqnarray}
A^{sm}_{ll'} \equiv 2\pi\int_a^b \text{d}x {}_s\lambda_{l'm} {}_s\lambda_{lm} =
\frac{2\pi}{(l+l'+1)(l-l')}&\biggl[&\left(x - \frac{s m}{ll'}\right)(l-l'){}_s
\lambda_{l'm}{}_s\lambda_{lm} \nonumber \\
&+& \frac{2l'+1}{C_{sl'm}} {}_s\lambda_{(l'-1)m} {}_s\lambda_{lm} -
\frac{2l+1}{C_{slm}} {}_s\lambda_{(l-1)m} {}_s\lambda_{l'm}\biggr]_a^b.
\end{eqnarray}
Note that $A_{ll'}^{sm} = A_{ll'}^{(-s)(-m)}$. The $l=l' \neq |m|$ integrals
can be evaluated recursively using
\begin{equation}
A^{sm}_{ll} = A^{sm}_{(l-1)(l-1)} 
+\frac{C_{slm}}{C_{s(l+1)m}}A^{sm}_{(l+1)(l-1)}
-\frac{C_{slm}}{C_{s(l-1)m}}A^{sm}_{l(l-2)}+\frac{2 s
m}{l(l^2-1)} C_{slm} A^{sm}_{l(l-1)},
\label{eq:appd1}
\end{equation}
which follows from Eq.~\eqref{Lambdarecur}. For $|m| > |s|$, the starting
values $A_{|m||m|}^{sm}$ can  be obtained from
\begin{eqnarray}
A^{sm}_{mm} &=& A^{s(m-1)}_{(m-1)(m-1)} + \frac{2\pi [x
{}_s\lambda_{mm}^2]_a^b}{2m+1}
+ \frac{s}{\sqrt{(2m+1)(m^2-s^2)}} A^{s(m-1)}_{m(m-1)} ,
\label{eq:appd2}
\\
A^{s(-m)}_{mm} &=& A^{s(-m+1)}_{(m-1)(m-1)} + \frac{2\pi [x
{}_s\lambda_{m(-m)}^2]_a^b}{2m+1} - \frac{s}{\sqrt{(2m+1)(m^2-s^2)}}
A^{s(-m+1)}_{m(m-1)} \qquad(m > |s|),
\label{eq:appd3}
\end{eqnarray}
which follow from the explicit form of ${}_s \lambda_{mm'}$ given in
Eq.~\eqref{eq:appc5}.
For $l'=l=m=|s|$ the integrals have the analytic solutions
\begin{equation}
A^{mm}_{mm} 
= \left[\left(\frac{x-1}{2}\right)^{2m+1} \right]_a^b,
\qquad A^{(-m)m}_{mm} = \left[\left(\frac{x+1}{2}\right)^{2m+1}\right]_a^b.
\end{equation}
For $|m| \leq |s|$, the starting values $A_{|s||s|}^{sm}$ for the recursion in
Eq.~\eqref{eq:appd1} can be obtained from the symmetry
$A_{ll'}^{sm} = A_{ll'}^{ms}$ and the recursion relations~\eqref{eq:appd2}
and \eqref{eq:appd3}.

If one is also generating the spin-weight zero overlap integrals for analysing
the temperature field an alternative approach is to
use Eq.~\eqref{inttheorem2} to relate the integrals of the spin-weight two
and zero harmonics. In
general integrals for different spin weights can be related by
\begin{eqnarray}
k_{l'(s-1)} k_{l's} \int_S \text{d}S\, {}_sY_{lm}^\ast\, {}_sY_{l'm'} &-& 
k_{l(s-1)} k_{ls}\int_S \text{d}S\,{}_{s-2}Y_{lm}^\ast\,{}_{s-2}Y_{l'm'}
\nonumber \\
&=& \oint_{\partial S}\deth \left( k_{l'(s-1)}\,{}_sY_{lm}^\ast
{}_{s-1}Y_{l'm'} + k_{ls}\,{}_{s-2}Y_{l'm'}{}_{s-1}Y_{lm}^\ast\right), 
\label{eq:appcdint}
\end{eqnarray}
where $k_{ls} \equiv \sqrt{l(l+1) - s(s-1)}$, and we have used the results
\begin{eqnarray}
\edth {}_s Y_{lm} &=&  k_{l(s+1)}\,\,{}_{s+1}Y_{lm}\\
\beth {}_{s}Y_{lm} &=& -k_{ls}\,\, {}_{s-1}Y_{lm} 
\end{eqnarray}
which follow from Eq.~\eqref{eq:appbharm}. For $s=0$ and $s=2$ one can
obtain the spin $\pm 2$ integrals in terms of the spin zero integrals. 
The spin zero integrals are computed using the above relations with $s=0$, in agreement with the relations given in Ref.~\cite{Wandelt00}.

For a small patch of sky a large number of the overlap integrals are
going to be very close to zero. This makes sense intuitively, and is easy
to see more quantitatively. From the differential
equation~\eqref{diffequation} the character of the harmonics changes
from oscillatory to decaying at the point where
\begin{equation}
\frac{m^2+s^2+2 m s x}{1-x^2} = l(l+1).
\end{equation}
For $x$ nearer the poles than the critical value the harmonics become
very small. For a sky patch extending from the north pole to $x$, for large $l$ we see that for $m^2$ larger than
$m^2\approx l^2(1-x^2)$ the harmonics will be small within the
patch. Hence the overlap integrals with $m^2 >
\text{min}(l^2,{l'}^2)(1-x^2)$ will be small, corresponding to one
(or both) of the harmonics being localized out of the sky
patch.

\section{Statistics of weak signal detection}
\label{App:Stats}

After obtaining a vector of observed data $\vBw$ two questions one might ask
are: (1) What is the probability that
the signal is just noise?; (2) Given the signal is not noise, what can
we say about the amplitude $r$? The likelihood function~\eqref{Likelihood}
encapsulates all the information in the data concerning the amplitude
$r$, and gives the posterior probability distribution of $r$ after
multiplying by the prior. Question (1) is really asking for a
comparison of two models, one in which $r=0$ (or $r<\epsilon$ where
$\epsilon$ is small), and one in which $r> 0$. Given there is a prior
probability $p$ that the signal is pure
noise ($r=0$) and probability $1-p$ that $r>0$, distributed with
the normalized prior probability distribution $f(r)$, the  posterior
probability that $r=0$ is given by
\begin{equation}
P(r=0|\vBw) = \frac{ p L(\vBw|r=0)}{ p L(\vBw|r=0) +
(1-p)\int_{0_+}^\infty \text{d}r L(\vBw|r) f(r)}.
\end{equation}
%
%
%
The posterior probability tells us the probability that the signal is
pure noise once we have a particular set of data. However, it doesn't
immediately tell us that we might expect to obtain from a given observation,
just what we know once the observation has been performed. The
posterior distribution can depend strongly on the prior.

The posterior probability is a monotonic function of the Bayes factor
\begin{equation}
t \equiv \frac{\int_{0_+}^\infty \text{d}r L(\vBw|r) f(r)}{L(\vBw|r=0)}.
\end{equation}
In classical hypothesis testing the likelihood ratio
$L(\vBw|r')/L(\vBw|r=0)$ is the most powerful test statistic for
distinguishing a model with $r=r'$ from one with $r=0$ (for details see
Ref.~\cite{kendall}).
The largest fraction $1-\alpha$ of the values of the likelihood ratio
under the null-hypothesis ($r=0$) determine a region, which, if the
observed ratio falls in it, rejects the null hypothesis at significance
$\alpha$.
The distribution of the likelihood ratio is straightforward to compute using
Monte-Carlo techniques. In general we do not have some fixed
alternative hypothesis $r=r'$; one possibility is to use the the statistic
$t$, formed by marginalizing over some prior, in place of the likelihood ratio.
Often the prior is fixed to $\delta(r-\hat{r})$, where $\hat{r}$ is
the maximum likelihood estimate of $r$, so the test reduces to a likelihood
ratio test with the alternative hypothesis $r=\hat{r}$.

By the time magnetic polarization comes to be
observed we should have some prior information about the tensor amplitude from
accurate measurement of the temperature power spectrum.  However at
this point we do not have very useful information about the prior
distribution, and to consider the possible results of future
experiments it is useful to assume no information.
The maximum entropy prior (the uniform prior) is probably not appropriate in
this case --- do we really think that all values of $r$ (below a certain
bound) are equally likely rather than, say, all values of $\ln r$
being equally likely? The
uniform prior gives radically different answers depending on the
choice of variable. The Jeffreys prior
\begin{equation}
f(r)\propto \left\la -\frac{\partial^2}{\partial r^2} \ln
L(r|\vBw)\right\ra^{1/2} = \half \Tr\left[
(\mN+r\mS)^{-1}\mS(\mN+r\mS)^{-1}\mS\right]^{1/2},
\end{equation}
is reparameterization invariant~\cite{kendall3}, and therefore does not
suffer from this problem. The Jeffreys prior
goes like $1/(1+r)$ if $\mN=\mS$ and in general is improper (does not have a
finite integral), though this is not a problem for evaluating test
statistics since the integral of the product with the likelihood
function will be finite. It is also not a true prior in the sense that
it depends on what data is going to be collected, but this is really a good
thing as it concentrates the prior probability in the region where we
need it in order to obtain a detection. Since the prior is not localized the
posterior probability of the null-hypothesis will depend on where one cuts
off the
prior. However if one only ever compares the Bayes factors $t$ the
cut-off is not very important since the likelihood function will be
localized. For this reason performing classical hypothesis tests using
$t$ is rather more independent of the prior information than
considering the values of the posterior probability, though clearly
it is the latter which is rigorous and contains all the available information.

Classical hypothesis tests are useful for assessing
probability of getting a detection at a given significance, though it
should be remembered that getting a detection at $99$ per cent confidence does
\emph{not} mean a the probability of $1$ per cent that the signal is pure
noise. However for sensible priors there will be a close
correspondence; classical and Bayesian techniques agree that high
values of $t$ correspond to the null-hypothesis being less
likely. Using classical hypothesis tests to compute the probability of getting
a detection at a given significance for a given true $r$ is significantly
simpler than computing and interpreting the corresponding distribution
of the posterior probability distributions.

For full sky surveys there are a large number of statistically-independent
magnetic variables, which can be thought of as the eigenvectors of
$r\mN^{-1/2}\mS\mN^{-1/2}$.
Those variables corresponding to large eigenvalues have high expected
signal to noise and are most useful
for obtaining detections. If $\vBw$ has dimension $n$ there will be
$n$ independent variables. If the eigenvalues of $r\mN^{-1/2}\mS\mN^{-1/2}$
are distributed fairly uniformly, and we consider expected chi-squared
detections of order one sigma, all the eigenvalues will be $\sim 1/\sqrt{n}$
and we have $r\mS \ll \mN$.
Using a second order approximation we then have
\begin{equation}
(\mN + r\mS)^{-1} \approx
(\mI - r\mN^{-1}\mS + r^2\mN^{-1}\mS\mN^{-1}\mS) \mN^{-1},
\end{equation}
and
\begin{equation}
\ln |\mN + r\mS| = \ln|\mN| + \ln|\mI +r\mN^{-1}\mS| \approx \ln|\mN| + \Tr(r\mN^{-1}\mS - \half r^2\mN^{-1}\mS\mN^{-1}\mS).
\end{equation}
In this approximation the entire likelihood distribution is simple to
compute since the 
matrix manipulations only need be performed once, and is
of the Gaussian form
\begin{equation}
L(\vBw|r) = L(\vBw|r=0) e^{\hat{r}^2/2\sigma^2}e^{-(r-\hat{r})^2/2\sigma^2},
\end{equation}
where 
\begin{equation}
1/\sigma^2 =  \vBw^\dag \mN^{-1}\mS\mN^{-1}\mS\mN^{-1}\vBw
- \half\Tr(\mN^{-1}\mS\mN^{-1}\mS),
\end{equation}
and the maximum likelihood estimate for $r$ is
\begin{equation}
\hat{r} \approx \half\sigma^2 \left[\vBw^\dag \mN^{-1} \mS
\mN^{-1} \vBw - \Tr(\mN^{-1}\mS)\right].
\end{equation}
The maximum likelihood estimate for $r$ is only weakly biased
for small signal to noise.  

Consider fixing the signal hypothesis to $r= \hat{r}$, which gives the
largest likelihood ratio possible for any prior.  The likelihood ratio
is then a monotonic function of the test statistic
\begin{equation}
\nu' = \hat{r}/\sigma \equiv \frac{\vBw^\dag \mN^{-1} \mS
\mN^{-1} \vBw - \Tr(\mN^{-1}\mS)}{\sqrt{4 \vBw^\dag \mN^{-1}\mS\mN^{-1}\mS\mN^{-1}\vBw
- 2\Tr(\mN^{-1}\mS\mN^{-1}\mS) }}.
\label{teststatistic}
\end{equation}
The quantity $\hat{r}/\sigma$ gives the number of standard deviations
the maximum likelihood is from pure noise, which is an good intuitive
measure of the number of `sigmas' at which the magnetic signal has
been detected. If any eigenvalues become large enough
then the small $r$ approximation will fail and one needs to compute the full
likelihood ratio to obtain optimal results.  The $\nu'$ statistic is a function only of
the vector $\mN^{-1/2}\vBw$ and the matrix
$\mN^{-1/2}\mS\mN^{-1/2}$, so working in the frame in which the matrix is
diagonal the statistic is simple to compute.  
Note that $\sigma^2$ and $\hat{r}$ can be negative
and correspond to a non-detection.

The second order approximation can also be used to speed up computing the
full likelihood function for integration against a prior. In the
diagonal frame it is straightforward to identify any modes with high signal to noise --- if there are some then the
likelihood for these modes can be computed exactly. The likelihood for
the remaining low signal to noise modes can be computed quickly using the 
approximation in which the likelihood is Gaussian. Multiplying these
together allows the full likelihood function to be computed, and hence the
 the probability distribution of $t$. Whilst this is slower than
using the $\nu'$ statistic it is a useful check, and may be essential
when there are very high signal to noise modes. 

For good detections the prior is not
very important as long as it is not small over the bulk of the
likelihood. If the prior is approximately constant over most of the
likelihood integral, the second order approximation is valid, and
$\hat{r}/\sigma\gg 0$, then $t \sim
e^{\hat{r}^2/2\sigma^2} g(\hat{r},\sigma)$ where $g(\hat{r},\sigma)$ depends weakly on $\hat{r}$
and $\sigma$ compared to the exponential. In this case $t$ depends
almost entirely (and monotonically) on the value of
$\nu'=\hat{r}/\sigma$, which is why $\nu'$ is a good statistic to
use. We found that
detection probabilities computed using full
likelihood results with the Jeffreys prior tend to agree very closely
with those computed using $\nu'$.

 The advantage of using $\nu'$ rather than
computing the full likelihood distribution is that it is
computationally significantly simpler and faster, which is useful,
though by no means essential, for
performing accurate Monte-Carlo computations. There is no problem
computing using the full likelihood distribution exactly from single samples of
actual observed data.

\subsection*{Busting the null-buster}

In the limit that $r\ll 1$ the likelihood function can be
approximated by the first two terms in its Taylor series in $r$. The
likelihood ratio for any given (small) $r$ is then a monotonic
function of $\left.\partial_r \ln L(\vBw|r)\right|_{r=0}$, which is
independent of $r$ and therefore provides the uniformly most powerful
test. For Gaussian signal and noise it is proportional to $2\hat{r}/\sigma^2$. After dividing by the
root of the variance in the null hypothesis $\la (2\hat{r}/\sigma^2)^2 \ra|_{r=0}= 2\Tr(\mN^{-1}\mS\mN^{-1}\mS)$ we obtain the quantity 
\begin{equation}
\nu \equiv \frac{ \vBw^\dag \mN^{-1} \mS \mN^{-1}\vBw
- \text{tr}(\mN^{-1}\mS)}{\sqrt{2\Tr(\mN^{-1}\mS\mN^{-1}\mS)}},
\end{equation}
which is the  optimal\footnote{It is the \emph{quadratic} statistic
which gives the maximal \emph{expected} detection in units of the
\emph{expected standard deviation} under the null hypothesis. None of these properties are
required or even especially desired (we are more interested in getting
detections at high \emph{significance} with high \emph{probability}).} quadratic `null-buster'
statistic introduced to the CMB literature in
Ref.~\cite{Tegmark98}. In the limit in which all the eigenvalues of $r\mN^{-1/2}\mS\mN^{-1/2}$ tend
to zero this is the
optimal test statistic. In general it is not --- a signal 
that can be detected will violate this assumption.  For a very large number of approximately equal
eigenvalues  the value of $\sigma$ will
approximate $\la 1/\sigma^2\ra^{-1/2}
=[\half\Tr(\mN^{-1}\mS\mN^{-1}\mS)]^{-1/2}$, the Fisher curvature for
small $r$. In this limit the null-buster remains a good
statistic and $\nu'/\nu \rightarrow 1$. However in general $\nu'$ performs
significantly better when the eigenvalues are distributed less evenly,
or when there are not that many eigenvalues. As the signal to noise increases
 $\nu'$ also performs much less sub-optimally than the null-buster.

For the case of magnetic polarization observations the $\nu'$ statistic outperforms the null-buster for a wide range of patch sizes in realistic
reionization models. The large scale magnetic signal
coming from low redshift ($z<10$) reionization gives a small
number of modes with relatively high signal to noise (see Fig.~\ref{contribs}), and the
conditions under which the null-buster is a good statistic are
therefore not satisfied. 
 The qualitative reason that the
null-buster performs poorly is that the position of the maximum and
the curvature of the
likelihood function are correlated, so dividing the actual maximum
by the expected curvature does not give you an accurate measure of
the number of `sigmas' from zero for a particular observation. This
makes the null hypothesis distribution unnecessarily broad at large values, and
therefore makes it harder to rule out the null hypothesis with good significance.
 The $\nu'$
statistic has a much sharper distribution than the null-buster (which
has a distribution similar to chi-squared) in the
alternative hypothesis, and the value of $\nu'$
corresponds much more closely to the significance (measured in
Gaussian-like `sigmas') of the result.

A few points can be made in the null-buster's defence. Firstly it is
slightly easier to compute than $\nu'$. Secondly, since we motivated the $\nu'$ by
assuming a Gaussian signal it is conceivable that the null-buster could perform better with certain
non-Gaussian signals. Lastly, the null-buster is quadratic which makes
it easy to calculate the mean and variance analytically. However it is
clear that with Gaussian signals using the null-buster is in general
significantly sub-optimal.



\end{document}